\documentclass[a4paper,11pt]{article}
\pdfoutput=1 

\usepackage{jheppub, amsmath, amssymb,amsthm,dsfont,graphicx,mathtools,multirow,rotating,placeins,verbatim} 
\usepackage[normalem]{ulem}

\usepackage{tensor}
\usepackage[T1]{fontenc} 
\usepackage[all,color]{xy}
\usepackage{float}

\newcommand{\sst}[1]{{\scriptscriptstyle #1}}

\def\0{{\sst{(0)}}}
\def\1{{\sst{(1)}}}
\def\2{{\sst{(2)}}}
\def\3{{\sst{(3)}}}
\def\4{{\sst{(4)}}}
\def\5{{\sst{(5)}}}
\def\6{{\sst{(6)}}}
\def\7{{\sst{(7)}}}

\newcommand{\be}{\begin{equation}}
\newcommand{\ee}{\end{equation}}
\def\ba{\begin{array}}
\def\ea{\end{array}}

\newcommand{\bea}{\begin{eqnarray}}
\newcommand{\eea}{\end{eqnarray}}

\DeclareMathOperator{\Tr}{Tr} 

\newcommand{\F}{\mathds{F}}

\def\be{\begin{equation}}
\def\ee{\end{equation}}
\def\ba{\begin{eqnarray}}
\def\ea{\end{eqnarray}}
\def\bi{\begin{itemize}}
\def\ei{\end{itemize}}

\def\F{\mathcal{F}}
\def\A{\mathcal{A}}
\def\zb{\bar{z}}

\def\Zb{\bar{Z}}
\def\L{\mathcal{L}}
\def\lamt{\tilde{\lambda}}
\def\Lamt{\tilde{\Lambda}}
\def\I{\mathcal{I}}
\def\delt{\tilde{\delta}}
\def\O{\mathcal{O}}
\def\deg{\text{deg}}
\def\Tr{\text{Tr}}

\title{A double copy for asymptotic symmetries in the self-dual sector}

\author[a]{Miguel Campiglia}
\author[b]{Silvia Nagy}

\affiliation[a]{Facultad  de  Ciencias, Universidad  de  la  Rep\'ublica\\
Ig\'ua  4225,  Montevideo,  Uruguay}

\affiliation[b]{Centre for Research in String Theory, School of Physics and Astronomy,\\
Queen Mary University of London, 327 Mile End Road, London E1 4NS, UK}

\emailAdd{campi@fisica.edu.uy}
\emailAdd{s.nagy@qmul.ac.uk}

\abstract{
We give a double copy construction for the symmetries of the self-dual sectors of Yang-Mills (YM) and gravity, in the light-cone formulation. We find an infinite set of double copy constructible symmetries. We focus on two families which correspond to the residual diffeomorphisms on the gravitational side. For the first one, we find novel non-perturbative double copy rules in the bulk. The second family has a more striking structure, as a non-perturbative gravitational symmetry  is obtained from a perturbatively defined symmetry on the YM side.   

At null infinity, we find the  YM origin of the subset of extended  Bondi-Metzner-Sachs (BMS) symmetries that preserve the self-duality condition. In particular, holomorphic large gauge YM symmetries are double copied to holomorphic supertranslations.  We also identify the single copy of superrotations with certain non-gauge YM transformations  that to our knowledge have not been previously presented in the literature. 

}

\begin{document} 
\maketitle
\flushbottom

\section{Introduction}
The idea of gravity as a double copy (DC)\footnote{See \cite{Bern:2019prr,Borsten:2020bgv,Borsten:2015pla} for some comprehensive reviews.} has found applications in the study of numerous aspects of the gravitational theory, most notably scattering amplitudes, where much of the recent success has been driven by the identification of a duality between the color and kinematic factors, the so-called Bern-Carrasco-Johannson (BCJ) duality \cite{Bern:2008qj,Bern:2010ue,Bern:2010yg,Bern:2017ucb,Bern:2012cd,Bern:2018jmv,BjerrumBohr:2012mg,Monteiro:2013rya}. It has progressed to the study of solutions, both perturbative \cite{Luna:2016hge,Goldberger:2016iau,Cardoso:2016ngt,Cardoso:2016amd,Luna:2020adi,Goldberger:2017ogt,Monteiro:2020plf} and exact \cite{Monteiro:2014cda,Ridgway:2015fdl,White:2016jzc,Luna:2018dpt,Berman:2018hwd,Bah:2019sda,Alfonsi:2020lub,Kim:2019jwm,Bahjat-Abbas:2020cyb,Lescano:2020nve,DeSmet:2017rve,Lee:2018gxc,Keeler:2020rcv,Alawadhi:2020jrv,Godazgar:2020zbv,White:2020sfn,Berman:2020xvs,Chacon:2021wbr}. Another important extension has been to find manifestations of it beyond gravity theory \cite{Carrasco:2016ygv,Carrasco:2016ldy,Cheung:2016prv,Borsten:2017jpt,Borsten:2019prq,Nagy:2014jza,Cachazo:2014xea}. A natural line of work following from these successes has been to formalise the correspondence by constructing double copy dictionaries for the fields themselves and by studying the symmetries of the theories, both local \cite{Anastasiou:2014qba,Anastasiou:2018rdx,LopesCardoso:2018xes,Alawadhi:2019urr,Banerjee:2019saj,Borsten:2021zir} and global \cite{Anastasiou:2017nsz,Borsten:2013bp,Anastasiou:2013hba,Anastasiou:2015vba,Anastasiou:2017taf}, as well as by giving direct constructions of Lagrangians via the DC \cite{Bern:2010yg,Tolotti:2013caa,Plefka:2018dpa,Plefka:2019wyg,Borsten:2020zgj,Borsten:2020xbt,Ferrero:2020vww}.  

Having a DC prescription for the local symmetries of the fields (i.e. the gauge transformations on the Yang-Mills (YM) side and diffeomorphisms on the gravity side) is of course a proof of the robustness of the DC dictionaries for the fields, but the possible applications of such a prescription go beyond this. For example, it was shown in \cite{Anastasiou:2018rdx,Borsten:2019prq,Borsten:2020xbt}, working in the BRST formalism, that one can exploit it to obtain a gauge-mapping algorithm, which gives a gravity gauge fixing functional output from a YM gauge-fixing functional input. This has been extended to the study of classical solutions in \cite{Luna:2020adi}. However, this has so far only been understood at linear level - going beyond this, either perturbatively (or non-perturbatively in certain special cases), could prove very powerful in the DC construction of classical solutions and/or Lagrangians.

A natural arena in which to discuss gravitational and gauge symmetries is  the asymptotic boundary of spacetime. In the case of (asymptotically) flat spacetimes, this is given by Penrose's null infinity \cite{Penrose:1962ij}. It has long been known that gauge and gravitational fields exhibit a rich structure at this boundary, with one of the earliest examples being the  Bondi-Metzner-Sachs (BMS) asymptotic symmetry group of gravity \cite{Bondi:1962px,Sachs:1962wk}. Whereas many of the  implications of these symmetries were known since the 80's thanks to Ashtekar and others \cite{Ashtekar:1981hw,Ashtekar:1981bq,Ashtekar:1987tt}, the subject experienced a veritable renaissance in recent years.  The trigger was twofold. On the one hand,  Barnich and Troessaert proposed an extension of BMS to include so-called superrotations \cite{Barnich:2009se}. On the other hand,  Strominger discovered  a link between future and past null infinity symmetries \cite{Strominger:2013jfa} which manifests as soft theorems in scattering amplitudes \cite{Strominger:2013lka,He:2014laa}. The ramifications of these developments are still ongoing, see  \cite{Strominger:2017zoo,Ashtekar:2018lor,Pasterski:2019ceq} for reviews.  

One may then ask whether the asymptotic perspective could shed light into the DC (and viceversa). In fact, there has already been  progress in this direction, through the use of DC for soft theorems  \cite{BjerrumBohr:2010hn,Oxburgh:2012zr,Naculich:2013xa,He:2014bga,White:2014qia,Vera:2014tda,Luna:2016idw,PV:2019uuv}, the memory effect \cite{Adamo:2017nia}, and in the context of celestial amplitudes \cite{Casali:2020vuy,Casali:2020uvr,Kalyanapuram:2020epb,Kalyanapuram:2020aya,Pasterski:2020pdk}. Additionally, a duality transformation relating the Schwarzschild and Taub-NUT solutions was shown to arise, via a DC construction, from an electric-magnetic duality relation \cite{Huang:2019cja,Emond:2020lwi}. Interestingly, this duality can be interpreted, on the gravity side, as the complexification of a particular supertranslation transformation.\footnote{It appears this duality could be the asymptotic version of the Ehlers transform discussed in \cite{Alawadhi:2019urr,Banerjee:2019saj}. It would be interesting to make this connection explicitly.  We thank Eoin O Colgain for bringing this possibility to our attention.} These promising results are calling for an interpretation from the perspective of asymptotic symmetries.

In this paper we present some progress in both directions outlined above. We will be restricting to the self-dual sector of gravity and YM theory, where the colour-kinematics duality was made manifest at the level of the algebra by Monteiro and O'Connell \cite{Monteiro:2011pc} (see also \cite{Monteiro:2013rya,Fu:2016plh,Cheung:2016prv,Chen:2019ywi,Elor:2020nqe,Chacon:2020fmr,Boels:2013bi} for related work and extensions), leading to a simple DC prescription for the construction of perturbative solutions. In light-cone gauge, the self-dual graviton and YM field have a natural description in terms of scalar fields $\phi$ and $\Phi$, respectively \cite{Plebanski:1975wn,Prasad:1979zc,Dolan:1983bp,Parkes:1992rz,Chalmers:1996rq,Cangemi:1996pf,Popov:1996uu,Popov:1998pc,Bardeen:1995gk}, and the self-duality conditions can be expressed as equations for these scalars, referred to as self-dual Einstein (SDE) and self-dual YM (SDYM) equations respectively. We will study the symmetries of these equations, with focus on those arising from the residual gauge symmetries of their respective parent theories. 
After laying down the `bulk' DC symmetry map (see below for details), we will move on to study  symmetries  at null infinity. On the gravitational side we will recover the subset of BMS that preserves the self-duality condition at null infinity. On the YM side, the single copy version of these symmetries will turn out to include large gauge  as well as certain non-gauge symmetries. The latter, to our knowledge, have not been discussed before in the asymptotic symmetry literature. 

The SDYM and SDE  equations are known to possess an infinite ladder of symmetries \cite{Popov:1996uu,Popov:1998pc,Dolan:1983bp,Prasad:1979zc,Park:1989vq,Husain:1993dp,Husain:1994vi,Wolf:2004hp,Popov:2006qu}, out of which the symmetries described above turn out to represent only  two rungs. We will see that the DC applies to all levels of this ladder. The asymptotic analysis of the higher level symmetries could in principle be worked out along the same lines as the ones studied here, but we do not attempt to do so in this work.

Let us now describe in more detail our proposed DC symmetry map. As anticipated above, we focus on two families of symmetries.  The first is exact in both YM and gravity, and constitutes a subset of diffeomorphisms on the gravity side and a subset of the non-abelian gauge symmetry on the YM side. We discover that a repackaging is required in order to manifest the color-kinematic duality and we find the appropriate way to perform the DC. Crucially, we find that the gauge parameter of YM is not mapped to the diffeomorphism parameter of gravity, but to a "Hamiltonian" of the diffeomorphism vector field w.r.t. the Poisson bracket inherent in the self-dual theory.
 Having performed this repackaging, the DC rules become remarkably simple. The second family is also exact on the gravity side, but perturbative and non-local on the YM side, presenting an apparent puzzle.  However, we find that the repackaging described above recasts the gravity transformation as a perturbative non-local series. This is then obtainable from its YM counterpart via the same simple rules as the first family, verifying the robustness of our construction. It is interesting to note that this second family is a subset of diffeomorphisms on the gravity side, but its single copy is not a subset of YM gauge transformations.   In the asymptotic limit, the first family maps holomorphic YM large gauge symmetries to holomorphic  supertranslations, whereas the second one identifies  certain (non-gauge and non-local) SDYM  transformations to holomorphic superrotations.\footnote{We have here  simplified the discussion in the interest of clarity. Asymptotically,  the second family splits into two subfamilies, see \autoref{fig:asymptsumm} for the complete picture.}

The paper is structured as follows:  In \autoref{Self-dual fields} we introduce our conventions and present a covariant description of self-dual fields in light-cone gauge. The corresponding residual gauge symmetries are discussed in \autoref{residualsec}. In \autoref{pertsymmsec} we describe a perturbative approach to general symmetries of the SDYM and SDE equations. In \autoref{Color to kinematics map for symmetries} we construct the DC map between the symmetries of these equations. Starting with a non-perturbative map for the first family of symmetries, we eventually extend its applicability to an infinite number of families.
 In \autoref{nullinfsec} we introduce coordinates adapted to null infinity and review gauge and gravity asymptotic symmetries. In \autoref{DCsymscri} we take the null-infinity limit of the first two families of symmetries described earlier and establish the DC map for asymptotic symmetries. We summarize our findings and discuss possible  future avenues in \autoref{Discussion}. Some of the technical discussions are given in appendices.

\section{Self-dual fields} \label{Self-dual fields}
Consider Minkowski spacetime written in light-cone coordinates $(U,V,Z,\Zb)$, related to Cartesian coordinates $X^\mu$ by:
\be
U=\frac{X^0-X^3}{\sqrt{2}}, \quad V=\frac{X^0+X^3}{\sqrt{2}}, \quad Z=\frac{X^1+i X^2}{\sqrt{2}}, \quad \Zb =\frac{X^1-i X^2}{\sqrt{2}}. \label{deflccoords}
\ee
We will split these light-cone coordinates in the following two sets of $2d$ coordinates:
\be
x^i :=(U,\Zb) , \quad y^{\alpha}:=(V,Z) \label{splitcoords}.
\ee
The spacetime metric  then takes the form
\be
ds^2= 2 \eta_{i \alpha} dx^i dy^\alpha = -2 dU d V+2 d Z d \Zb .
\ee
The description of self-dual fields and their symmetries will involve the ``area element'' of these 2d spaces, described by 
\ba
 \Omega_{i j} d x^i d x^j & = & d U d \Zb - d \Zb d U \\ 
  \Pi_{\alpha \beta}  d y^\alpha d y^\beta &= & d V d Z - d Z d V
\ea
We will treat $\Omega_{i j}$ and $ \Pi_{\alpha \beta}$ as antisymmetric 4 dimensional tensors with $\Omega_{\alpha \mu}$ and $ \Pi_{i\mu}$ vanishing. By raising indices with the inverse spacetime metric $\eta^{i \alpha}$, we can regard these two tensors as (partial) inverses of each other:
\ba
\Omega_i^{\ \alpha} \Pi_{\alpha}^{\ j} = \delta_i^{j} \label{OmegaPi} \\
\Pi_{\alpha}^{\ i} \Omega_{i}^{\ \beta} = \delta_\alpha^{\beta} \label{PiOmega}.
\ea
The self-dual YM  and metric fields will be described in terms of scalar fields $\Phi$ and $\phi$ according to
\ba
\mathcal{A}_\mu & =&  \Pi_\mu^{\ \nu} \partial_\nu \Phi \label{Aphimu} \\
h_{\mu \nu} & =& \Pi_\mu^{\ \rho} \Pi_\nu^{\ \sigma} \partial_\rho \partial_\sigma \phi \label{hphimu}.
\ea
In the notation of \eqref{splitcoords}, the non-zero components take the form
\ba
\mathcal{A}_\alpha & =&  \Pi_\alpha^{\ i} \partial_i \Phi \label{Aphi} \\
h_{\alpha \beta} & =& \Pi_\alpha^{\ i} \Pi_\beta^{\ j} \partial_i \partial_j \phi \label{hphi},
\ea
with the scalar fields satisfying \footnote{See \autoref{sdapp} for the derivation of these equations from the self-duality conditions for YM and gravity.}  
\ba
\square \Phi & = & - i \Pi^{i j}[\partial_i \Phi,\partial_j \Phi]  \label{eomPhi} \\
\square \phi & = & \frac{1}{2}\Pi^{i j} \Pi^{k l} \partial_i \partial_k \phi \partial_j \partial_l \phi \label{eomphi}
\ea
where $\square  \equiv 2 \eta^{i \alpha} \partial_i \partial_\alpha$ is the wave operator. We will refer to  \eqref{eomPhi} and \eqref{eomphi} as the self-dual YM (SDYM) and self-dual Einstein (SDE) equation respectively. We are using conventions where the coupling constants are absorbed in the field, so that the field strength is $\F_{\mu \nu} = \partial_\mu \A_\nu  - \partial_\nu \A_\mu - i [\A_\mu,\A_\mu]$,  the spacetime metric $g_{\mu \nu}= \eta_{\mu \nu}+h_{\mu \nu}$, and there is no explicit coupling constant in the field equations.

In \cite{Monteiro:2011pc}, the notion of a kinematic algebra in the self-dual sector was defined as a Poisson bracket algebra. In our conventions, the Poisson bracket (PB)  is defined as:   
\be
 \{f,g\}:= \Pi^{ij}\partial_i f \partial_j g
\ee
One can now recover \eqref{eomphi} from\eqref{eomPhi} via the replacement rules \cite{Monteiro:2011pc}:
\be  \label{color_kin_eom}
-i[ \ , \ ] \to \frac{1}{2} \{ \ , \ \}. \quad \Phi \to \phi.
\ee
 Note that our conventions are slightly different from \cite{Monteiro:2011pc}, and so are the color to kinematics replacement rules.

\subsection{Symmetries of the self-dual field equations} \label{deltodeltsec}
The self-dual YM and Einstein equations have long been known to possess infinitely many symmetries \cite{Popov:1996uu,Popov:1998pc,Dolan:1983bp,Prasad:1979zc,Park:1989vq,Husain:1993dp,Husain:1994vi}. These can be constructed recursively as follows. Let $\delta \Phi$ and $\delta \phi$ be symmetries of the SDYM and SDE equations respectively:
\ba
\square \delta \Phi & = & - 2 i \Pi^{i j}[\partial_i \Phi,\partial_j \delta \Phi]  \label{eomdelPhi} \\
\square \delta \phi & = & \Pi^{i j} \Pi^{k l} \partial_i \partial_k \phi \partial_j \partial_l  \delta \phi \label{eomdelphi}
\ea
One can then obtain new symmetries $\delt \Phi$ and $\delt \phi$, defined implicitly by the condition\footnote{See e.g. Eq. (15) of \cite{Cangemi:1996pf} for the YM case.}
\ba
\partial_i \delt \Phi & = & \Omega_i^{\ \alpha} \partial_\alpha \delta \Phi - i [\partial_i \Phi, \delta \Phi] \label{defdeltPhi}  \\
\partial_i \delt \phi & =&  \Omega_i^{\ \alpha} \partial_\alpha \delta \phi + \frac{1}{2} \{\partial_i \phi, \delta \phi\} \label{defdeltphi}
\ea
In each case, one can verify the consistency condition  $\partial_{[i} \partial_{j]} \delt=0$ is satisfied. By direct computation one can then check that $\delt \Phi$ and $\delt \phi$ satisfy \eqref{eomdelPhi} and \eqref{eomdelphi} provided the respective SDYM and SDE equations are satisfied. We will refer to this $\delta \to \delt$ map as the \emph{symmetry raising map} on each respective theory. It will turn out to be useful for identifying the DC rules for asymptotic symmetries.

We conclude the section by noting that  the scalar fields also have  intrinsic redundancies:  The field $\Phi$ should be thought of as defined modulo an arbitrary function $\Psi(y)$:
\be
\Phi \sim \Phi +\Psi(y) \label{redunPhi}
\ee
as this does not affect the YM field (\ref{Aphi}). Similarly, $\phi$ is defined modulo
\be
\phi \sim \phi + \psi^{(0)}(y) + x^i \psi^{(1)}_i (y). \label{redunphi}
\ee
which preserves  (\ref{hphi}).\footnote{The attentive reader might notice that whereas \eqref{redunPhi} is a symmetry of the field equation (\ref{eomPhi}), this does not appear to be the case for \eqref{redunphi} and  \eqref{eomphi}. The reason for this is that the actual self-dual condition on the metric, which  is of course invariant under \eqref{redunphi}, takes the form of a differential operator acting on \eqref{eomphi} - see \autoref{App Metric field}. }  Note that (\ref{redunPhi}) and (\ref{redunphi}) are not the standard gauge symmetries of the YM and metric fields (to be discussed in the next section). Rather, they are additional redundancies that arise in the description given by  (\ref{Aphi}) and (\ref{hphi}).

\section{Residual gauge symmetries} \label{residualsec}
In this section we describe the gauge transformations on the YM and gravity fields that preserve the form (\ref{Aphi}), (\ref{hphi}). It is interesting to note that these fields satisfy the following gauge conditions
\ba \label{harmgge}
\partial^\mu \mathcal{A}_\mu =0, \quad \partial^\mu h_{\mu \nu}=0, \quad \eta^{\mu \nu}h_{\mu \nu }=0.
\ea
Thus, the symmetries that we  will be  discussing can be thought of as a subset of the residual gauge symmetries of conditions \eqref{harmgge}.


\subsection{YM field}

We start with the YM field. Given a gauge parameter $\Lambda$, the corresponding gauge transformation is
\be
\delta_\Lambda \mathcal{A}_\mu = \partial_\mu \Lambda + i [\Lambda, \mathcal{A}_\mu]. \label{delA}
\ee
If we want to preserve the condition $\A_i=0$, we must have
\be
\partial_i \Lambda =0 \implies \Lambda=\Lambda(y) \label{lam}
\ee
We now ask if a gauge transformation with such gauge parameter is compatible with (\ref{Aphi}). That is, we ask whether there is a $\delta_\Lambda \Phi$ such that
\be
\delta_\Lambda \mathcal{A}_\alpha  = \Pi_\alpha^{\ i} \partial_i \delta_\Lambda \Phi \label{delA2}.
\ee
By substituting (\ref{Aphi}) in (\ref{delA}),  the LHS of (\ref{delA2}) can be written as 
\ba
\delta_\Lambda \mathcal{A}_\alpha & = & \partial_\alpha \Lambda +i    [\Lambda, \Pi_\alpha^{\ i} \partial_i \Phi]  \\
&=& \partial_\alpha \Lambda + i  \Pi_\alpha^{\ i} \partial_i  [\Lambda,  \Phi], \label{delA3}
\ea
where to get the last line we used (\ref{lam}). The second term in (\ref{delA3}) is of the required form (\ref{delA2}).  We find it convenient to use the notation
\be
\delta_\Lambda \Phi = \delta^{(0)}_\Lambda \Phi + \delta^{(1)}_\Lambda \Phi
\ee
where $\delta^{(0)}_\Lambda \Phi$ is independent of $\Phi$ and $\delta^{(1)}_\Lambda \Phi$ is linear in $\Phi$. We emphasize that this is an exact symmetry and that we are not doing a perturbative expansion at this stage. Eq. (\ref{delA3}) then tells us that
\be
 \delta^{(1)}_\Lambda \Phi = i  [\Lambda,  \Phi] .
\ee
It now remains to see whether the  first term in (\ref{delA3}) can also be written in the form (\ref{delA2}). That is, we would like to find  $\delta^{(0)}_\Lambda \Phi$ such that
\be
 \Pi_\alpha^{\ i} \partial_i  \delta^{(0)}_\Lambda \Phi = \partial_\alpha \Lambda  \label{delA0}.
\ee
Using (\ref{OmegaPi}), this equation can equivalently be written as:
\be
 \partial_i  \delta^{(0)}_\Lambda \Phi =  \Omega_i^{\ \alpha} \partial_\alpha \Lambda.
\ee
Since the RHS is independent of $x^i$, we can integrate it directly to obtain\footnote{We ignore an integration `constant' $\Psi(y)$ due to the redundancy (\ref{redunPhi}).}
\be
\delta^{(0)}_\Lambda \Phi =   \Omega_i^{\ \alpha} x^i \partial_\alpha \Lambda.  \label{del0lamphi}
\ee
Summarizing: The gauge symmetries compatible with (\ref{Aphi}) are parametrized by gauge parameters $\Lambda(y)$, and act on $\Phi$ according to
\be
\delta_\Lambda \Phi = \Omega_i^{\ \alpha} x^i \partial_\alpha \Lambda +  i  [\Lambda,  \Phi]. \label{dellamPhi}
\ee
Additionally, one could also consider symmetries of the scalar $\Phi$ which do not arise as YM gauge transformations, but are intrinsic symmetries of the equation \eqref{eomPhi}. As we will show in \autoref{Color to kinematics map for symmetries}, one of these families of transformations can be obtained as the single copy of a gravitational diffeomorphism symmetry.

\subsection{Metric field}
In the gravitational case, the spacetime metric is given by 
\be 
g_{\mu \nu}= \eta_{\mu \nu} + h_{\mu \nu}. 
\ee
Let us emphasize this is not just a first order metric but the full-non linear (self-dual) metric. In particular, the inverse metric is exactly given by  $g^{\mu \nu}= \eta^{\mu \nu} - h^{\mu \nu}$  since $\eta^{\rho \sigma}h_{\mu \rho} h_{\sigma \nu}=0$. Except for the metric tensor $g$, we will be raising and lowering indices with $\eta$.

Given a vector field $\xi^\mu$, the gauge transformation on $h_{\mu \nu}$ is defined from the lie derivative on $g_{\mu \nu}$:
\be
  \delta_\xi h_{\mu \nu} :=  \L_\xi \eta_{\mu \nu} + \L_\xi h_{\mu \nu} \label{delh}
\ee
We will be looking for vector fields that are independent of $h$, so we can study each term in (\ref{delh}) separately. We start by imposing the condition that the first term in (\ref{delh}) preserves $h_{i \mu}=0$:
\be
\L_\xi \eta_{i \mu} = 2 \partial_{(i} \xi_{\mu)}=0,
\ee
where $\xi_\mu \equiv \eta_{\mu \nu} \xi^\nu$, with $\xi^\nu$ the infinitesimal diffeomorphism vector field. 
Setting $\mu=i,\alpha$ we obtain
\be
2 \partial_{(i} \xi_{j)}=0 \implies \xi_i = a_{i j}(y) x^j - c_i(y) \quad \text{with $a_{(ij)}=0$} ,
\ee
 and
\be\label{xi1}
\partial_i \xi_\alpha + \partial_\alpha  \xi_i =0 \implies  \partial_i \xi_\alpha = - \partial_\alpha  a_{i j} x^j+ \partial_\alpha c_i
\ee
An equation of the form $\partial_i \xi_\alpha = v_{i \alpha}$ can be solved only if $\partial_{[i}v_{j] \alpha}=0$. This implies
\be
\partial_\alpha a_{ij}=0.
\ee
A constant $a_{ij}$ represents a rigid rotation in the $x^i$ plane and is hence part of the global Poincare symmetries. Since we are interested in local symmetries, we set $a_{ij}=0$.  Integrating (\ref{xi1})  we are led to
\be
\xi_\alpha =  \partial_\alpha c_i(y) x^i + b_\alpha(y)
\ee
with $b_\alpha(y)$ an integration ``constant''.  To continue, we need to (i) determine if the second term in (\ref{delh}) is compatible with $h_{i \mu}=0$ and (\ref{hphi}) and (ii) determine if $2\partial_{(\alpha} \xi_{\beta)}$ can be written in a form compatible with (\ref{hphi}). To simplify the analysis, we will discuss  separately the cases  $c_i=0$ and $b_\alpha=0$. The general case can be obtained as a linear combination of these two simple cases.
\subsubsection{First family of symmetries} \label{firstvfsec}
Setting $c_i=0$, we have 
\be
\xi_{i}=0, \quad \xi_\alpha=b_\alpha(y). \label{xiizero}
\ee
and so the resulting vector field is given by\footnote{Recall we raise and lower indices with $\eta$. For the vector field (\ref{xialphazero}) this happens to be equivalent to raising and lowering with $g$, but this will not be always the case.}
\be
\xi^i = \eta^{i \alpha} b_\alpha(y), \quad \xi^\alpha =0 .  \label{xialphazero}
\ee
Let us now discuss the second term in (\ref{delh})
\be
\L_\xi h_{\mu \nu} = \xi^\rho \partial_\rho h_{\mu \nu} + 2 \partial_{(\mu}\xi^{\rho}h_{\nu) \rho}. \label{lieh}
\ee
From (\ref{xialphazero}) and $h_{i \mu}=0$ we find that the second term in (\ref{lieh}) vanishes. The first term clearly respects $h_{i \mu}=0$. Furthermore, using Eq. (\ref{hphi}) it can be written as
\ba
\xi^k \partial_k h_{\alpha \beta} &= &\xi^k \partial_k  \Pi_\alpha^{\ i} \Pi_\beta^{\ j} \partial_i \partial_j  \phi   \\
 &= &  \Pi_\alpha^{\ i} \Pi_\beta^{\ j} \partial_i \partial_j  (\xi^k \partial_k \phi )\label{del1h1}
\ea
where in the second line we used the fact that $\partial_i \xi^k=0$. Eq (\ref{del1h1}) gives the part  of $\delta_\xi \phi$ that is linear in $\phi$:
\be
\delta^{(1)}_\xi \phi = \xi^k \partial_k \phi. \label{del1phi1}
\ee
It  remains to see whether  the first term in (\ref{delh}) is compatible with (\ref{hphi}). Thus,  we ask if there is a $\delta^{(0)}_\xi \phi$ such that
\be
 \Pi_\alpha^{\ i} \Pi_\beta^{\ j} \partial_i \partial_j  \delta^{(0)}_\xi \phi = 2\partial_{(\alpha} \xi_{\beta)}.
\ee
Using (\ref{OmegaPi}) and (\ref{xiizero}) this equation can be written as:
\be
 \partial_i \partial_j  \delta^{(0)}_\xi \phi = 2  \Omega_i^{\ \alpha}  \Omega_j^{\ \beta} \partial_{(\alpha} b_{\beta)}.
\ee
Since $b_\beta$ is independent of the $x^i$ variables, the equation  can be directly integrated, leading to\footnote{As in the YM case, we do not include the `integration constants' that are part of the ambiguity in the definition of $\phi$ (\ref{redunphi}).}
\be
\delta^{(0)}_\xi \phi  =  \Omega_i^{\ \alpha}  \Omega_j^{\ \beta}  x^i x^j  \partial_{\alpha} b_{\beta}. \label{del0xi1}
\ee
The  family of vector fields (\ref{xialphazero}) then acts on the scalar field according to
\be
\delta_\xi \phi =  \Omega_i^{\ \alpha}  \Omega_j^{\ \beta}  x^i x^j  \partial_{\alpha} b_{\beta}  + \eta^{i \alpha} b_\alpha \partial_i \phi. \label{delxiphi1}
\ee
At this stage we notice that $\delta_\xi \phi$ does not appear to be a symmetry of the SDE equation \eqref{eomphi}. To see what has gone wrong, we need to recall (see  \autoref{sdapp}) that the SDE equation for $\phi$ (\ref{eomphi}) is a sufficient but not necessary condition to produce a self-dual metric. The actual self-duality condition takes the form of a 2nd order differential operator acting on the SDE equation (\ref{eomphi}), see Eqs. \eqref{sdgrav1} and \eqref{sdgrav2}. Of course the self-duality condition is invariant under the transformation \eqref{delxiphi1}, since it arises from an infinitesimal diffeomorphism. 
In order to have a symmetry of the SDE equation (\ref{eomphi}), one can check that it is sufficient to restrict $b_\alpha(y)$ to be a total derivative, 
\be
b_\alpha(y) = \partial_\alpha b(y) \label{bpartialb}.
\ee
We will refer to the resulting symmetry transformation as $\delta_b$, with:
\be
\delta_b \phi =  \Omega_i^{\ \alpha}  \Omega_j^{\ \beta}  x^i x^j  \partial_{\alpha} \partial_\beta b 
+ \eta^{i \alpha} \partial_\alpha b \partial_i \phi. \label{delxiphi1partial}
\ee
We will later see that (\ref{bpartialb}) does not actually represent any restriction on the asymptotic vector fields at null infinity.

\subsubsection{Second family of symmetries} \label{2ndvfsec}
Setting $b_\alpha=0$ we have 
\be
\xi_i = - c_i(y) , \quad \xi_\alpha =   \partial_\alpha c_j(y) x^j . \label{xic}
\ee
We now start by demanding the first term in (\ref{delh}) to be compatible with (\ref{hphi}), as this will impose restrictions on $c_i$: 
\be
2\partial_{(\alpha} \xi_{\beta)} = 2  \partial_\alpha \partial_\beta c_k(y) x^k =\Pi_\alpha^{\ i} \Pi_\beta^{\ j} \partial_i \partial_j  \delta^{(0)}_\xi \phi 
\ee
for some $\delta^{(0)}_\xi \phi$. Multiplying the equation with $\Omega$'s as before, the last equality is equivalent to 
\be
  \partial_i \partial_j  \delta^{(0)}_\xi \phi  = 2  \Omega_{i}^{\ \alpha} \Omega_{j}^{\ \beta}\partial_\alpha \partial_\beta c_k(y) x^k =:c_{ijk}x^k.
\ee
For this to be integrable, we need the tensor $c_{ijk}$ to be symmetric under the exchange of any two indices. The expression as given is already symmetric in $(ij)$. One can show that, in order to make it symmetric in $(jk)$ (and hence in $(ik)$) one needs to take
\be
c_k(y) = \Omega_{k}^{\ \gamma}\partial_\gamma c(y)
\ee
for some function $c(y)$. The resulting expression can then be integrated, leading to 
\be
\delta^{(0)}_\xi \phi = \frac{1}{3} \Omega_{i}^{\ \alpha} \Omega_{j}^{\ \beta}\Omega_{k}^{\ \gamma}x^i x^j x^k \partial_\alpha \partial_\beta \partial_\gamma c \label{delxiphiinhom}.
\ee
We finally discuss the second term in (\ref{delh})
for the vector field $\xi^{\mu}$:
\be
\xi^i  =    \eta^{i \alpha }\Omega_{j}^{\ \beta} x^j \partial_\alpha \partial_\beta c  ,\quad 
\xi^\alpha = -  \Omega^{\alpha \beta} \partial_\beta c \label{xicvector}.
\ee
One can show that such a vector field preserves the condition $h_{i \mu}=0$ as well as the form (\ref{hphi}). To simplest way to show this is to verify the vector field leaves the tensor $\Pi_\mu^{\ \nu}$ invariant
\be
\L_{\xi} \Pi_\mu^{\ \nu} = 0,
\ee
from which it follows that $\L_\xi h_{i \mu}=0$ and
\be
 \L_\xi h_{\alpha \beta} =  \frac{1}{4}\Pi_\alpha^{\ i} \Pi_\beta^{\ j} \partial_i \partial_j  \delta^{(1)}_\xi \phi  \label{liexih2}
\ee
where 
\be
\delta^{(1)}_\xi \phi = \L_\xi \phi = \xi^i \partial_i \phi + \xi^\alpha \partial_\alpha \phi . \label{del1phi2}
\ee 
To summarize, the second family of vector fields acts on $\phi$ as: $\delta_\xi \phi= \delta^{(0)}_\xi \phi+\delta^{(1)}_\xi \phi$, with $\delta^{(0)}_\xi \phi$ and $\delta^{(1)}_\xi \phi$ given by (\ref{delxiphiinhom}) and  (\ref{del1phi2}) respectively.

 We emphasize that both families are exact, non-perturbative symmetries of the field equations (\ref{eomphi}). As we will  discuss in  \autoref{DCanddeltodeltsec}, it turns out that the second family can be obtained from the first one by the symmetry raising map given in Eq. (\ref{defdeltphi}).

\section{Perturbative approach to symmetries} \label{pertsymmsec}
In this section we present a perturbative approach to symmetries of the SDYM and SDE equations. This will offer an alternative perspective on the symmetries described before. The perturbative approach will be useful when constructing the DC symmetry map in \autoref{Color to kinematics map for symmetries}. We will present expressions up to first order in the fields. The all-order expansions are given recursively in \autoref{Recursive construction at arbitrary orders}.

\subsection{YM field}
The field equations (\ref{eomPhi}) may be cast as a tower of field equations associated to a perturbative expansion\footnote{Recall we have absorbed the coupling constant in the definition of the field; the order in the expansion counts the number of $\Phi^{(0)}$ fields.}
\be
\Phi = \Phi^{(0)}+ \Phi^{(1)}+ \cdots
\ee
such that
\ba
\square \Phi^{(0)} & = & 0 \label{eomPhi0} \\
\square \Phi^{(1)} & = & -  i \Pi^{i j}[\partial_i \Phi^{(0)},\partial_j  \Phi^{(0)}] \\
\cdots 
\ea
We now look at symmetries $\delta \Phi$ of the field equations (\ref{eomPhi}) in the above perturbative scheme. That is, we look for
\be
\delta \Phi = \delta \Phi^{(0)}+  \delta \Phi^{(1)} + \cdots
\ee
such that
\ba
\square \delta \Phi^{(0)} & = & 0  \label{eomdelPhi0}\\
\square \delta \Phi^{(1)} & = & -  2 i \Pi^{i j}[\partial_i \Phi^{(0)},\partial_j  \delta \Phi^{(0)}] \label{eomdelPhi1} \\
\cdots 
\ea
We may think of these equations as determining $\delta \Phi$ in terms of a ``seed'' $\delta \Phi^{(0)}$: Given $\delta \Phi^{(0)}$ satisfying (\ref{eomdelPhi0}), $\delta \Phi^{(1)}$ is defined implicitly by (\ref{eomdelPhi1}), and similarly for the higher order terms. Generically this would lead to a $\delta \Phi^{(1)}$ that depends non-locally on $\Phi^{(0)}$ due to the need to invert the wave operator. The case of gauge symmetries discussed previously is an exception. From (\ref{del0lamphi}) we see that in this case the seed is given by
\be
\delta \Phi^{(0)} =   \Omega_j^{\ \beta} x^j \partial_\beta \Lambda(y),  \label{delphi0}
\ee
which satisfies (\ref{eomdelPhi0}):
\be
\square \delta \Phi^{(0)} = 2 \eta^{i \alpha}\partial_i \partial_\alpha \delta \Phi^{(0)} = 2 \Omega^{\alpha \beta}\partial_\alpha \partial_\beta \Lambda =0.
\ee
If we now substitute (\ref{delphi0}) in the RHS of (\ref{eomdelPhi1}) we find
\ba
 -  2 i \Pi^{i j}[\partial_i \Phi^{(0)},\partial_j  \delta \Phi^{(0)}] & = &   -  2 i \eta^{i \beta} [\partial_i \Phi^{(0)},\partial_\beta \Lambda]\\ 
& = & - i \square [\Phi^{(0)},\Lambda]  \label{boxPhi0Lam}
\ea
where in the first equality we used $\Pi^{i j}  \Omega_j^{\ \beta} = \eta^{i \beta}$ and in the second equality we used (\ref{eomPhi0}) and the fact that $\Lambda$ is independent of $x^i$. From (\ref{eomdelPhi1}) we then conclude that
\be
\delta \Phi^{(1)} = - i  [\Phi^{(0)},\Lambda], \label{delta_1_Phi_pert} 
\ee
consistent with the result from the previous section.   One could continue and obtain the higher order terms, which in this case will resum to the exact symmetry (\ref{dellamPhi}). See \autoref{perturbativeapp} for further discussion on higher order terms.

\subsection{Metric field}
Proceeding analogously for the ``metric'' field $\phi$, we have
\be
\phi = \phi^{(0)}+ \phi^{(1)}+ \cdots,
\ee
\ba
\square \phi^{(0)} & = & 0 \label{eomphi0} \\
\square \phi^{(1)} & = & \frac{1}{2}\Pi^{i j} \Pi^{k l} \partial_i \partial_k \phi^{(0)} \partial_j \partial_l \phi^{(0)} \\
\cdots 
\ea
A symmetry $\delta \phi = \delta \phi^{(0)}+  \delta \phi^{(1)} + \cdots$ of the field equations will then satisfy
\ba
\square \delta \phi^{(0)} & = & 0  \label{eomdelphi0}\\
\square \delta \phi^{(1)} & = & \Pi^{i j} \Pi^{k l} \partial_i \partial_k \phi^{(0)} \partial_j \partial_l \delta \phi^{(0)} \label{eomdelphi1} \\
\cdots 
\ea
We now discuss how the residual gauge symmetries found in the previous section fit into this description.

\subsubsection{First family of symmetries}
For the first family of vector fields described earlier, Eq. (\ref{delxiphi1partial}) gives the following candidate for the ``seed'' $\delta \phi^{(0)}$:
\be
\delta \phi^{(0)}  =  \Omega_i^{\ \alpha}  \Omega_j^{\ \beta}  x^i x^j  \partial_{\alpha} \partial_\beta b \label{del0phi1b},
\ee
which satisfies $\square \delta \phi^{(0)}=0$. If we now substitute (\ref{del0phi1b}) in the RHS of (\ref{eomdelphi1}) we obtain
\ba
\Pi^{i j} \Pi^{k l} \partial_i \partial_k \phi^{(0)} \partial_j \partial_l \delta \phi^{(0)} & = & 2 \eta^{k \beta} \eta^{i \alpha}\partial_i \partial_k \phi^{(0)}  \partial_{\alpha} \partial_\beta b  \\
& = & \square  \eta^{i \alpha} \partial_i  \phi^{(0)}  \partial_{\alpha} b
\ea
where to get the last equality we used (\ref{eomdelphi0}) and the fact that $b$ is independent of $x^i$. Comparing with (\ref{eomdelphi1}) we conclude
\be
 \delta \phi^{(1)} =  \eta^{i \alpha} \partial_i  \phi^{(0)}  \partial_{\alpha} b,
\ee
compatible with the expected result from \eqref{delxiphi1partial}.

\subsubsection{Second family of  symmetries}
For the second family of  symmetries, we take $\delta \phi^{(0)}$ to be given by (\ref{delxiphiinhom})
\be
\delta \phi^{(0)} = \frac{1}{3}  \Omega_{i}^{\ \alpha} \Omega_{j}^{\ \beta}\Omega_{k}^{\ \gamma} x^i x^j x^k\partial_\alpha \partial_\beta \partial_\gamma c \label{del0phi2},
\ee
It can be easily checked that the above satisfies (\ref{eomdelphi0}). Substituting in the RHS of (\ref{eomdelphi1}) one finds
\be
\Pi^{i j} \Pi^{k l} \partial_i \partial_k \phi^{(0)} \partial_j \partial_l \delta \phi^{(0)}  = - 2  \eta^{i \beta} \eta^{l \gamma}  \Omega_m^{\ \alpha} x^m \partial_i \partial_l \phi^{(0)} \partial_\alpha \partial_\beta \partial_\gamma c . \label{rhodelphi2}
\ee
Since it is not immediately obvious how to ``pull out'' a d'Alambertian from such an expression, let us take guidance from the previous section. From (\ref{del1phi2}) and (\ref{xicvector}) we expect to have
\be
\delta \phi^{(1)}=  \eta^{i \alpha} \Omega_j^{\ \beta}x^j \partial_i \phi^{(0)} \partial_\alpha \partial_\beta c- \Omega^{\alpha \beta}\partial_\alpha  \phi^{(0)} \partial_\beta c. \label{del1phi2p}
\ee
Indeed, one can check by direct computation that the d'Alambertian of (\ref{del1phi2p}) coincides with (\ref{rhodelphi2}),  upon using (\ref{eomphi0}).

\section{Color to kinematics map for symmetries}\label{Color to kinematics map for symmetries}
According to \cite{Monteiro:2011pc}, the color to kinematics map for the self-dual sector consists in replacing YM matrix commutators by certain  Poisson brackets on the $x^i$ variables, as reviewed in \eqref{color_kin_eom}: 
\be \label{color_kin_5_1}
\Phi \to \phi,\quad  -i[ \ , \ ] \to \tfrac{1}{2} \{ \ , \ \}.
\ee
This will be our starting point in understanding the double copy for symmetries. In the following section, we will make the color-kinematics duality manifest at the level of the symmetry transformations. A key ingredient in this will be to recast the gravitational symmetry in a form that is analogous to the YM one. We will achieve this by expressing the gravitational symmetry parameter in terms of a Hamiltonian $\lambda$ (w.r.t. the Poisson bracket) associated to the diffeomorphism vector field. This will allow us to supplement the rules \eqref{color_kin_5_1} with  
\be
\Lambda \to \lambda 
\ee
where $\Lambda$ is the YM symmetry parameter and $\lambda$ is the Hamiltonian describing the gravitational symmetry.  

\subsection{First family of symmetries}\label{First family of symmetries}
We will start by comparing the gauge symmetries found in the YM case (\ref{dellamPhi})  with the first family of symmetries found in the gravitational case \eqref{delxiphi1partial}:
\ba
\delta_\Lambda \Phi & = & \Omega_i^{\ \alpha} x^i \partial_\alpha \Lambda +  i  [\Lambda,  \Phi]  \label{delLamPhi} \\
\delta_b \phi &= &   \Omega_i^{\ \alpha}  \Omega_j^{\ \beta}  x^i x^j  \partial_{\alpha} \partial_\beta b + \eta^{i \alpha}   \partial_{\alpha} b \partial_i  \phi. \label{delbphi}
\ea
Let us first focus on the terms that are linear in the scalar fields. According to \eqref{color_kin_5_1}, the YM term linear in $\Phi$ should be mapped to 
\be
-i [\Lambda, \Phi] \to \tfrac{1}{2}\{\lambda,\phi \}
\ee
for some $\lambda$.  In order to reproduce the linear term in $\delta_b \phi$ we take
\be
\lambda = 2 \Omega_i^{\ \alpha} x^i \partial_\alpha b, \label{lamitob}
\ee
which can be thought of as the ``Hamiltonian'' of the vector field $\xi^i = \eta^{i \alpha} \partial_\alpha b$ with respect to the Poisson bracket $\tfrac{1}{2}\{ , \}$.

We now notice that the $\phi$-independent piece of $\delta_b \phi$ can be written in terms of $\lambda$, leading to an expression that has the same form as the $\Phi$-independent piece of $\delta_\Lambda \Phi$. Thus, we can express the total $\delta_b \phi$ in terms of $\lambda$ as:
\be\label{delta_b_grav}
\delta_b \phi  =  \tfrac{1}{2}\Omega_i^{\ \alpha} x^i \partial_\alpha \lambda -\tfrac{1}{2}\{\lambda,\phi \}. 
\ee
Comparing the expression above with \eqref{delLamPhi}, we find that the tranformation rules map into each other under 
\be \label{full_rules}
\Phi \to \phi,\quad  -i[ \ , \ ] \to \tfrac{1}{2} \{ \ , \ \},\quad \Lambda \to \lambda
\ee
with an apparent mismatch of $\tfrac{1}{2}$ for the inhomogeneous terms. However we note that this is resolved by expressing the transformation rules on the YM and gravitational fields themselves, as: 
\be
\begin{aligned}
\delta_\Lambda \A_\alpha  =&  \partial_\alpha\Lambda +  i \Pi_\alpha^{\ i}\partial_i [\Lambda,  \Phi]  \\
\delta_\lambda h_{\alpha\beta}  =&  \Pi_{(\alpha}^{\ i}\partial_i \left( \partial_{\beta)} \lambda - \tfrac{1}{2}\Pi_{\beta)}^{\ j}\partial_j\{\lambda,\phi \}\right)
\end{aligned}
\ee 
If we insist on working at the level of the scalar fields, then the replacement rules need to be augmented by the inclusion of a  multiplicative factor for the inhomogeneous term:
\be
\mathfrak{r}=\frac{\text{deg}(\Lambda)+1}{\text{deg}(\lambda)+1}  \label{defn}
\ee
where $\text{deg}(f)$ counts the power of $x^i$ in a function $f$ which is homogeneous in the $x^i$ variables. 
 This numerical factor dissapears in the replacement rules for
\be
\A_\alpha=\Pi_\alpha^{\ i}\partial_i\Phi \to h_{\alpha\beta}=\Pi_\alpha^{\ i}\Pi_\beta^{\ j}\partial_i\partial_j\phi, 
\ee
due to the action of the derivatives w.r.t $x^i$ on the inhomogeneous term.  \\
Finally, we note that
\be
\text{deg}(\lambda)=\text{deg}(\Lambda)+1  
\ee
and moreover
\be\label{Lambda_explicit}
\Lambda(y) \to \lambda= 2 \Omega_i^{\ \alpha} x^i \partial_\alpha b(y).
\ee

\subsection{Second family of symmetries}\label{Second family of symmetries}
Having understood how the first family of gravitational symmetries can be obtained from the YM ones, we now focus on the second family of gravitational symmetries:
\be
\delta_c \phi = \frac{1}{3}  \Omega_{i}^{\ \alpha} \Omega_{j}^{\ \beta}\Omega_{k}^{\ \gamma} x^i x^j x^k\partial_\alpha \partial_\beta \partial_\gamma c  +  \xi^i \partial_i \phi + \xi^\alpha \partial_\alpha \phi \label{deltildec}
\ee
where
\be
\xi^i= \eta^{i \alpha} \Omega_j^{\ \beta}x^j  \partial_\alpha \partial_\beta c, \quad \xi^\alpha=- \Omega^{\alpha \beta} \partial_\beta c.
\ee
We first note that the vector field $\xi^i$ can again be written in terms of a `Hamiltonian'
\be
\tilde{\lambda}=  \Omega_i^{\ \alpha} \Omega_j^{\ \beta} x^i x^j  \partial_\alpha \partial_\beta c \label{lamtilde}
\ee
so that $\xi^i \partial_i \phi = \tfrac{1}{2}\{\phi,\tilde{\lambda} \}$. Using (\ref{lamtilde}),  we can rewrite (\ref{deltildec}) as 
\be
\delta_c \phi = \frac{1}{3}  \Omega_{i}^{\ \alpha}  x^i \partial_\alpha \lamt  - \tfrac{1}{2}\{ \lamt, \phi \} - \Omega^{\alpha \beta} \partial_\alpha \phi \partial_\beta c. \label{delcphi}
\ee
This brings $\delta_c \phi$ into a form that resembles the one obtained before for $\delta_b \phi$ in \eqref{delta_b_grav}, with the exception of the final term. To understand this extra term we will  use the perturbative description of symmetries given in \autoref{pertsymmsec}, as described below.

We would like to find a symmetry of the SDYM theory such that, upon the appropriate color to kinematics replacement rules, one recovers the second family of gravitational symmetries \eqref{delcphi}. Guided by \eqref{Lambda_explicit}, we find that the natural choice for the $\Phi$-independent piece of this  YM symmetry is
\be
\delta^{(0)}_{\Lamt} \Phi = \frac{1}{2}  \Omega_{i}^{\ \alpha}  x^i \partial_\alpha \Lamt \label{delta0Lamt}
\ee
with $\Lamt$ of the form 
\be
\Lamt = \Omega_{i}^{\ \alpha}x^i \partial_\alpha B(y)
\ee
for some function $B(y)$. This generalizes the $\Phi$-independent part of the symmetry (\ref{delLamPhi}) to the case where $\Lambda$ is linear in $x$ . Restricting to the inhomogeneous terms in the transformations for the YM field and the graviton,
\be
\begin{aligned}
\delta_{\Lamt}^{(0)} \A_\alpha  =&  \partial_\alpha\Lamt   \\
\delta_{\lamt}^{(0)} h_{\alpha\beta}  =&  \Pi_{(\alpha}^{\ i}\partial_i  \partial_{\beta)} \lamt \label{dellamth}
\end{aligned}
\ee 
we find that they are related by the same rules \eqref{full_rules} as the first family.\footnote{\label{footnote_not_gauge}We emphasize here that $\delta_{\Lamt}$ is not defined directly on $\A_\mu$, but through its action on the scalar $\Phi$. In particular, using \eqref{Aphimu}, we find $\delta^{(0)}_{\Lamt} \A_i\equiv\Pi_i^{\ \mu}\partial_\mu\delta^{(0)}_{\Lamt}\Phi =0$, due the fact that $\Pi_i^{\ \mu}$ vanishes identically. Note that $\delta_{\Lamt}^{(0)} \A_i\neq\partial_i\tilde{\Lambda}$, so it cannot be interpreted as a gauge transformation, even at the lowest order.} Alternatively, one can work directly at the level of the scalar fields by taking into acount the multiplicative factor \eqref{defn}, which in this case is $\mathfrak{r}=2/3$, using $\text{deg}(\tilde{\Lambda})=1$ and $\text{deg}(\tilde{\lambda})=2$.

The idea now is to use the perturbative  method of \autoref{pertsymmsec} to obtain the $\O(\Phi^1)$ symmetry transformation generated by (\ref{delta0Lamt}). We will then discuss how the color to kinematic rules applied to this term reproduce the two  $\mathcal{O}(\phi^{1})$ terms in (\ref{delcphi}).

\subsubsection{$\mathcal{O}(\Phi^1)$ symmetry}
Following the perturbative method of \autoref{pertsymmsec}, we would like to solve for $\delta_{\Lamt} \Phi^{(1)}$ in Eq. (\ref{eomdelPhi1}),
\be
\square \delta_{\Lamt} \Phi^{(1)}  =  -  2 i \Pi^{i j}[\partial_i \Phi^{(0)},\partial_j  \delta_{\Lamt} \Phi^{(0)}] \label{eomdelPhiLamt}
\ee
 given the `seed'
\be
\delta_{\Lamt} \Phi^{(0)} = \frac{1}{2}  \Omega_{i}^{\ \alpha}  x^i \partial_\alpha \Lamt \label{deltaLamtPhi0}.
\ee
We proceed in a similar fashion as in the analysis between Eqs. (\ref{delphi0}) and (\ref{boxPhi0Lam})  where we verified the perturbative method against the  residual YM gauge symmetry.  Noting that $\partial_j  \delta_{\Lamt} \Phi^{(0)} = \Omega_{j}^{\ \alpha}  \partial_\alpha \Lamt  $, the RHS of (\ref{eomdelPhiLamt}) can be written 
\ba
 -  2 i \Pi^{i j}[\partial_i \Phi^{(0)},\partial_j  \delta_{\Lamt} \Phi^{(0)}] & = & -2 i \eta^{i \alpha}[\partial_i \Phi^{(0)},\partial_\alpha \Lamt] \\
 & = & - i \square [ \Phi^{(0)}, \Lamt]  + 2 i \eta^{i \alpha}[\partial_\alpha \Phi^{(0)},\partial_i \Lamt]
\ea
where in the last line we used that $\square \Phi^{(0)} =0 = \square \Lamt$. Unlike for the residual YM symmetry, there is now a reminder term when one `pulls out' the wave operator. We then conclude that
\be
 \delta_{\Lamt} \Phi^{(1)}  = - i  [ \Phi^{(0)}, \Lamt]  + 2 i \square^{-1} \eta^{i \alpha}  [\partial_\alpha \Phi^{(0)},\partial_i \Lamt] \label{delLamtPhi1} .
\ee

\subsubsection{$\O(\phi^1)$ symmetry from the DC}
If we apply the color to kinematic map
\be
-i[ \ , \ ] \to \tfrac{1}{2} \{ \ , \ \},  \quad \Phi \to \phi , \quad \Lamt \to \lamt \label{ctok2nd}
\ee
to expression (\ref{delLamtPhi1}) one finds
\be
\delta_{\Lamt} \Phi^{(1)}  \to  \tfrac{1}{2}  \{ \phi^{(0)}, \lamt \}  -  \square^{-1} \eta^{i \alpha} \{ \partial_\alpha \phi^{(0)}, \partial_i \lamt \}  \label{ctokdelLamtPhi1}
\ee
The first term corresponds to the second term in \eqref{delcphi}. For the second term we have
\ba
  - \square^{-1} \eta^{i \alpha} \{ \partial_\alpha \phi^{(0)}, \partial_i \lamt \} & = & - \square^{-1} \eta^{i \alpha} \Pi^{jk}  \partial_j \partial_\alpha \phi^{(0)} \partial_k \partial_i \lamt \\
& = &  - 2\square^{-1} \eta^{i \alpha} \Pi^{jk}  \partial_j \partial_\alpha \phi^{(0)}  \Omega_i^{\ \beta} \Omega_k^{\ \gamma}   \partial_\beta \partial_\gamma c \\
& = &  - 2 \square^{-1} \Omega^{\alpha \beta} \eta^{j \gamma}  \partial_j \partial_\alpha \phi^{(0)}     \partial_\beta \partial_\gamma c \\
 & = &  - 2 \square^{-1} \Omega^{\alpha \beta} \eta^{j \gamma}  \partial_j \partial_\gamma (\partial_\alpha \phi^{(0)}     \partial_\beta  c) \\
 & = & - \Omega^{\alpha \beta}\partial_\alpha \phi^{(0)}     \partial_\beta  c , \label{finalctokdel1philamt}
\ea 
where we used the definition of the Poisson bracket, the expression for $\lamt$ as given in \eqref{lamtilde}, and the fact that $\square \phi^{(0)}$ vanishes. Expression (\ref{finalctokdel1philamt})  reproduces, in the perturbative setting, the last term in  \eqref{delcphi}. This establishes $\delta_{\Lamt} \Phi^{(1)}$ as the inverse DC map of $\delta_{c} \phi^{(1)}$. In fact, as shown in \autoref{perturbativeapp} the DC map can be extended to arbitrary order in the perturbative  expansion.  

\subsection{Summary}
Let us summarize our findings. By associating certain ``Hamiltonians'' to the infinitesimal residual diffeomorphisms of the SDE equations, we managed to obtain simple DC rules mapping SDYM to SDE symmetries. In order to compactly write these rules for the scalar fields, let us introduce the notation
\be
S := \Omega_i^{\ \alpha}  x^i \partial_\alpha \label{defS}.
\ee
The first family of symmetries take the form
\ba
\delta_\Lambda \Phi & = & S(\Lambda) +  i  [\Lambda,  \Phi] \label{delLamsummary} \\
\delta_\lambda \phi &= &   \tfrac{1}{2}S(\lambda) -\tfrac{1}{2}\{\lambda,\phi \}, \label{dellamsummary}
\ea
where on the YM side we have a gauge transformation with parameter $\Lambda$ independent of $x^i$, and on the gravitational side we have an infinitesimal diffeomorphism with `Hamiltonian'  $\lambda$ that is linear in $x^i$. The two are related by the DC map \eqref{full_rules}, supplemented by the multiplicative coefficient $\mathfrak{r}$  in the inhomogeneous term (see discussion around \eqref{defn} for details):
\be \label{full_rules_summary}
\Phi \to \phi,\quad  -i[ \ , \ ] \to \tfrac{1}{2} \{ \ , \ \},\quad \Lambda \to \lambda,\quad S \to \mathfrak{r} S, \quad 
\mathfrak{r}=\frac{\text{deg}(\Lambda)+1}{\text{deg}(\lambda)+1}.
\ee
The second family can be written as
\ba
\delta_{\Lamt} \Phi & = & \tfrac{1}{2} S(\Lamt) - i  [ \Phi, \Lamt]  + 2 i \square^{-1} \eta^{i \alpha}  [\partial_\alpha \Phi,\partial_i \Lamt] +\O(\Phi^2) \label{delLamtsummary} \\
\delta_{\lamt} \phi &= & \tfrac{1}{3}  S(\lamt)  +\tfrac{1}{2}\{  \phi,\lamt \}  -  \square^{-1} \eta^{i \alpha} \{ \partial_\alpha \phi, \partial_i \lamt \}  +\O(\phi^2) \label{dellamtsummary}
\ea
The two are related by exactly the same DC rules \eqref{full_rules_summary} as for the first family. Although here we have presented the argument up to linear order in the field, the map can in fact be extended to arbitrary order, see \autoref{perturbativeapp}. Let us emphasize that the original gravitational symmetry, as given in \eqref{delcphi}, is non-perturbative. In \eqref{dellamtsummary}, the higher order terms arise  as a result of parametrizing this symmetry in terms of the Hamiltonian $\lamt$. This requires the use of the field equations, which introduce higher order terms. 

The second family displays two interesting features. Firstly, it maps a perturbative expression on the YM side to a non-perturbative one on the gravity side, as described above. Secondly, while on the gravity side the symmetry is a diffeomorphism, on the YM side we have a non-gauge symmetry of the field equation \eqref{eomPhi}.

\subsection{Relation with the symmetry raising map} \label{DCanddeltodeltsec}
In \autoref{deltodeltsec} we described the  existence of  symmetry raising maps on YM and gravity self-dual theories that, given a symmetry $\delta$, produce a new symmetry $\delt$ according to Eqs. \eqref{defdeltPhi}, \eqref{defdeltphi}. 

Since, as described in \autoref{pertsymmsec} the symmetries we are studying are fully determined by their inhomogenous piece, it is natural to restrict these equations to the inhomogenous parts of the symmetries:
\ba
\partial_i \delt^{(0)} \Phi & = & \Omega_i^{\ \alpha} \partial_\alpha \delta \Phi^{(0)}  \label{delt0Phi}  \\
\partial_i \delt^{(0)} \phi & =&  \Omega_i^{\ \alpha} \partial_\alpha \delta \phi^{(0)} . \label{delt0phi}
\ea
It is now straightforward to check that, if we take $\delta \Phi^{(0)} = \delta_\Lambda \Phi^{(0)}$ and $\delta \phi^{(0)} = \delta_\lambda \phi^{(0)}$,  the resulting $\delt^{(0)}$ are given by $\delta^{(0)}_{\Lamt} \Phi$ and $\delta^{(0)}_{\lamt} \phi$ with
\ba
\Lamt &  = &  \delta^{(0)}_{\Lambda} \Phi =  S( \Lambda)   \label{summ_2nd_rec} \\
\lamt & = & \delta^{(0)}_{\lambda} \phi = \tfrac{1}{2}  S( \lambda)   .\label{summ_2nd_rec_grav}
\ea
Since the symmetries are fully determined by their inhomogenous piece, the above shows that the map $\delta \to \delt$ takes the full $\delta_{\Lambda} \Phi$  into the full $\delta_{\Lamt} \Phi$ (and similarly for $\delta_\lambda \phi$). We can summarize the above structure by the diagram
\be\label{diag1}
\begin{array}{cccc}
 \delta_\Lambda \Phi & \xrightarrow{\text{DC}} & \delta_\lambda \phi  \\
 \downarrow & & \downarrow \\
\ \delta_{\Lamt} \Phi  & \xrightarrow{\text{DC}} & \delta_{\lamt} \phi,
\end{array}
\ee
where the horizontal arrows represent the DC map between each family of symmetries, and the vertical arrows the $\delta \to \delt$ map on each theory. 

Said in different words, we have verified that the symmetry raising map commutes with the DC symmetry map. We will use this result in the next section to obtain the DC symmetry map at null infinity for the second family of symmetries.

\subsection{Double copy for an infinite family of symmetries}\label{Double copy for an infinite family of symmetries}
We found that the second family of symmetries arises from applying the raising map on the first family. It is clear that one can continue this process indefinitely and construct an infinite tower of symmetries on both sides. A natural question is whether the DC will continue to hold at each level. In other words, we want to check whether the diagram \eqref{diag1} extends indefinitely:  
\be\label{diag}
\begin{array}{cccc}
 \delta_{1} \Phi & \xrightarrow{\text{DC}} & \delta_{1} \phi  \\
 \downarrow & & \downarrow \\
 \delta_{2} \Phi  & \xrightarrow{\text{DC}} & \delta_{2} \phi\\
 \downarrow & & \downarrow \\
 \vdots & & \vdots \\
 \delta_{n} \Phi  & \xrightarrow{\text{DC}} & \delta_{n} \phi\\
 \vdots & & \vdots
\end{array}
\ee
where $\delta_n$ denotes the symmetry of the $n$-th family. The first two lines of the diagram correspond to the original diagram \eqref{diag1}. It is easy to show by recursion that the DC holds at every level. We have explicitly demonstrated this for $\delta_1$ and $\delta_2$ in \autoref{First family of symmetries}, \autoref{Second family of symmetries}, and \autoref{perturbativeapp}. Assuming that the DC holds at level $n-1$, it immediately follows from the definition of the $n$-th level symmetry \eqref{defdeltPhi} and \eqref{defdeltphi}:  
\ba
\partial_i \delta_n \Phi & = & \Omega_i^{\ \alpha} \partial_\alpha \delta_{n-1} \Phi - i [\partial_i \Phi, \delta_{n-1} \Phi]  \label{summ_recursive_YM}\\
\partial_i \delta_n \phi & =&  \Omega_i^{\ \alpha} \partial_\alpha \delta_{n-1} \phi + \frac{1}{2} \{\partial_i \phi, \delta_{n-1} \phi\} \label{summ_recursive_grav}
\ea
that $\delta_n \phi$ is obtained from $\delta_n \Phi$ via the DC rules \eqref{full_rules_summary}. The focus in this paper is on the first two families, since they are the ones corresponding to asymptotic diffeomorphisms on the gravitational side.  It is however instructive to display the general form of the inhomogenous $n$-th transformations. Consider first the YM case, starting with the `zeroth'  $n=0$ family:
\be
\delta_0 \Phi =\Lambda (y)   
\ee
Notice that these are just the redundancies described in \eqref{redunPhi}. It is then easy to solve \eqref{summ_recursive_YM} at $\mathcal{O}(\Phi^0)$. One finds
\be
\delta_n^{(0)} \Phi =\frac{1}{n!} S^{n}(\Lambda)   .
\ee
As we did for the second family in \eqref{summ_2nd_rec}, we can introduce a gauge-like parameter 
\be
\Lambda_n=\delta_{n-1}^{(0)} \Phi =\frac{1}{(n-1)!}S^{n-1}(\Lambda)
\ee
 at  $n$-th level, such that
\be  \label{deln0Phi}
\delta_n^{(0)} \Phi =\frac{1}{n} S(\Lambda_n)  = \frac{1}{\deg(\Lambda_n)+1} S(\Lambda_n) .
\ee
We refer to $\Lambda_n$ as a gauge-like parameter since
\be
\delta_n^{(0)}\mathcal{A}_\alpha  =  \Pi_\alpha^{\ i} \partial_i \delta_n^{(0)}\Phi=\partial_\alpha \Lambda_n  . \label{deln0A}
\ee
Note however that $\delta_n^{(0)}\mathcal{A}_i=0 \neq \partial_i  \Lambda_n $ (see \autoref{footnote_not_gauge} for $n=2$). Let us now discuss the gravitational case. Recall that in this case there are two kind of redundancies \eqref{redunphi}. In order to be consistent with the diagram \eqref{diag}, one is lead to start the recursion at $n=-1$ with,
\be
\delta_{-1} \phi =2 b(y).
\ee
Solving  \eqref{summ_recursive_grav} at $\mathcal{O}(\phi^0)$. One finds
\be
\delta_n^{(0)} \phi =\frac{1}{(n+1)!} S^{n+1}(2 b) =   \frac{1}{(n+1)!} S^{n}(\lambda) , \quad \lambda := S(2 b) ,
\ee
where we expressed the result in terms of the first family  Hamiltonian $\lambda$.
The $n=-1,0$ levels  describe the two  redundancies in \eqref{redunphi}. As we did for the second family \eqref{summ_2nd_rec_grav}, we can define a  `Hamiltonian' parameter 
\be
\lambda_n=\delta_{n-1}^{(0)} \phi =\frac{1}{n!}S^{n-1}(\lambda)
\ee
 at  $n$-th level, such that
\be \label{deln0phi}
\delta_n^{(0)} \phi =\frac{1}{n+1} S(\lambda_n)  = \frac{1}{\deg(\lambda_n)+1} S(\lambda_n) . 
\ee
As in the first two families, with this definition of $\lambda_n$ one has
\be
\delta_{n}^{(0)} h_{\alpha\beta}  =  \Pi_{(\alpha}^{\ i}\partial_i  \partial_{\beta)} \lambda_n, \label{deln0h}
\ee
corresponding  with the  DC version of \eqref{deln0A}.  Comparing \eqref{deln0Phi} with \eqref{deln0phi} we verify these are indeed compatible with \eqref{full_rules_summary}. The DC map can be extended to arbitrary order in the fields, following \autoref{Recursive construction at arbitrary orders}. This extends the applicability of the rules \eqref{full_rules_summary} to all $n$ levels,
\be \label{DCrulesinhom}
\Phi \to \phi,\quad  -i[ \ , \ ] \to \tfrac{1}{2} \{ \ , \ \},\quad \Lambda_n \to \lambda_n,\quad S \to \mathfrak{r} S, \quad 
\mathfrak{r}=\frac{\deg(\Lambda_n)+1}{\deg(\lambda_n)+1}.
\ee\\

\noindent \emph{Comment:} It may naively appear that there exists a different DC prescription, one that maps the $n$-th YM family to the $(n-1)$-th gravity one.  Upon relabeling  the gravity subscripts,   this would yield identical zeroth order expressions on both sides. However this prescription has various drawbacks. Firstly, it would map YM gauge symmetries $\delta_1 \A_\mu = \delta_\Lambda \A_\mu=\partial_\mu \Lambda + i [\Lambda, \mathcal{A}_\mu]$ into trivial diffeomorphisms $\delta_0 h_{\mu \nu}=0$. In particular, the large YM gauge symmetries at null infinity would not have a gravitational counterpart. Secondly, the second family of YM symmetries would be mapped to the first family of gravitational ones. As we shall see, this would imply an additional mismatch in the number of would-be DC related asymptotic symmetries.
Finally, our original DC prescription \eqref{diag1} is shown in \autoref{convoapp} to be consistent with the convolution DC \cite{Anastasiou:2014qba,Anastasiou:2018rdx,LopesCardoso:2018xes} where they overlap, whereas the alternative DC would not be.


\section{Large gauge symmetries at null infinity} \label{nullinfsec}
\subsection{Asymptotic fields at null infinity}
In this section we switch to coordinates adapted to null infinity. Given the light cone coordinates $(U,V,Z,\Zb)$ introduced in \eqref{deflccoords}, a natural choice for Bondi-type coordinates $(r,u,z,\zb)$ is given by:
\be
U= r z \zb + u, \quad V= r, \quad Z=r z , \quad \Zb = r \zb, \label{defbondicoords}
\ee
in terms of which the Minkowski line element takes the form
\be
ds^2 = -2 du dr + 2 r^2 d z d \zb.
\ee
Future null infinity $\I$ is reached by taking $r \to \infty$ while keeping the rest of the variables fixed. It is thus parametrized by  $(u,z,\zb)$.  Note that we  will be working in a flat conformal frame  for which the  (degenerate) metric at null infinity is  $2 dz d \zb$.\footnote{This can be contrasted with the often used Bondi frame for which the 2d metric is that of the unit sphere. See e.g. appendix A of \cite{Kapec:2017gsg} for a comparison between the two.} 

Massless fields are captured at null infinity by certain $r \to \infty$ leading components. For a massless scalar field, this is given by
\be
\varphi(r,u,z,\zb) \stackrel{r \to \infty}{=} \varphi_{\I}(u,z,\zb)/r + \cdots \label{scalarfallr}
\ee
where the dots indicate terms that decay faster than $1/r$. $\varphi_{\I}(u,z,\zb)$ is to be regarded as a field defined intrinsically at null infinity, which contains all the radiative data of $\varphi$. We will also refer to it as the `free data' at $\I$, since it may be regarded as the analogue of the Cauchy data for the `final time' hypersurface $\I$, see e.g. \cite{Ashtekar:1981bq}.  Similarly, gauge and metric fields at null infinity are captured by
\ba
\A_z(r,u,z,\zb) & \stackrel{r \to \infty}{=} & A_z(u,z,\zb) +  \cdots \label{Afallr} \\
h_{zz}(r,u,z,\zb) & \stackrel{r \to \infty}{=} &  r C_{zz}(u,z,\zb) +  \cdots \label{hfallr}
\ea
and respective $z \leftrightarrow \zb$ expressions. If we now consider expressions (\ref{Aphi}) and (\ref{hphi}) in Bondi coordinates \eqref{defbondicoords} we find
\ba
\A_z = r \partial_u \Phi, \quad  \A_{\zb} = 0, \\
h_{zz}= r^2 \partial^2_u \phi, \quad h_{\zb \zb}=0,
\ea
from which one easily finds how the gauge and gravity free data is encoded in the scalar fields free data (see \eqref{scalarfallr})  
\be
A_z= \partial_u \Phi_{\I}, \quad   A_{\zb}  = 0 \label{AzitoPhi}
\ee
\be
C_{zz} = \partial^2_u \phi_{\I}, \quad   C_{\zb \zb}  = 0. \label{Czzitophi}
\ee
We thus recover the expected result that self-dual fields are associated with data at null infinity that has vanishing antiholomorphic components.

In the reminder of the section we will study how the `bulk' symmetries described in the previous sections manifest at null infinity,  with the aim of obtaining DC rules for asymptotic symmetries.  Before getting started, however, it will be useful to briefly review the standard description of asymptotic symmetries.
\subsection{Review of large gauge symmetries at null infinity} \label{Review of large gauge symmetries at null infinity} 
In the general (not necessarily self-dual) case, one considers gauge and diffeomorphism symmetries that are non-trivial at null infinity. In the YM case, these are generated by gauge parameters that asymptote to a non-trivial function on the celestial sphere \cite{Strominger:2013lka,He:2015zea},
\be
\Lambda(r,u,z,\zb) \stackrel{r \to \infty}{=} \Lambda_0(z,\zb) + \cdots
\ee
where the dots denote terms that go to zero as $r \to \infty$. The induced symmetry transformation on the null infinity radiative data is then
\be
\delta_{\Lambda_0} A_z = \partial_z \Lambda_0 + i [\Lambda_0, A_z]. \label{delLambdaAz}
\ee
In the gravitational case one finds two classes of vector fields: The Bondi-Metzner-Sachs supertranslations \cite{Bondi:1962px,Sachs:1962wk}, parametrized by functions $f(z,\zb)$,
\be
\xi_f = f(z,\zb) \partial_u + \cdots
\ee
and the Barnich-Troessaert superrotations \cite{Barnich:2009se}, parametrized by holomorphic 2d vector fields $Y^{z}(z) \partial_z+Y^{\zb}(\zb) \partial_{\zb}$,
\be
\xi_Y  =   Y^z(z)\partial_z + \frac{u}{2}\partial_z Y(z) \partial_u - \frac{1}{2} \partial_z Y(z) r \partial_r + (z \leftrightarrow \zb) + \cdots   ,
\ee
where in both cases the dots denote terms that vanish when $r \to \infty$. The infinitesimal diffeomorphisms generated by these vector fields induce the following action on the null infinity radiative data:
\ba
\delta_f C_{zz} & = & -2 \partial^2_z f + f \partial_u C_{zz} \label{delfCzz} \\
\delta_Y C_{zz}& =& - u \partial^3_z Y^z +  (Y^z \partial_z + 2 \partial_z Y^z +\frac{u}{2} \partial \cdot Y \partial_u -\frac{1}{2}\partial \cdot Y) C_{zz} \label{delYCzz}
\ea 
where $\partial \cdot Y = \partial_z Y^z + \partial_{\zb}Y^{\zb}$. The transformation properties for $C_{\zb \zb}$ can be obtained by doing the replacement $z \to \zb$ in the expressions  above. 

Whereas the above represent the simplest large gauge symmetries, it is possible to find additional ones by carefully relaxing the standard fall-off conditions \eqref{Afallr}, \eqref{hfallr}. In particular, one can make sense of gauge symmetries with $O(r)$ gauge parameter \cite{Campiglia:2016hvg,Laddha:2017vfh,He:2019pll,Campiglia:2021oqz} and non-holomorphic superrotations \cite{Campiglia:2015yka,Compere:2018ylh,Campiglia:2020qvc}. Interestingly, the $O(r)$ gauge symmetries will turn out to play a key role in identifying the single copy version of superrotations, see  \autoref{DCsymscri}.

Coming back to the self-dual case,  we can ask what is the subset of the previously discussed symmetries that preserve the conditions 
\be
A_{\zb}=0 , \quad C_{\zb\zb}=0
\ee 
at null infinity. On the YM side, this condition is preserved provided
$\Lambda_0$ is holomorphic,
\be
\Lambda_0= \Lambda_0(z).
\ee
On the gravity side, the supertranslation function has to be of the form
\be
f = f(z) + \zb g(z).
\ee
Finally, the antiholomorphic component of the superrotation vector field has to be of the form $Y^{\zb}= a_0 + a_1 \zb + a_2 \zb^2$ for some constants $a_0$, $a_1$ and $a_2$. This corresponds to Lorentz generators and is hence part of the global symmetry group. If, as in \autoref{residualsec}, we only keep track of local symmetries, we are left with vector fields of the form
\be
Y^{z} = Y(z), \quad Y^{\zb}=0. \label{Yzbzero}
\ee
 Below we will recover these restricted symmetries as the null infinity limit of the residual gauge symmetries discussed in  \autoref{residualsec}.

\subsection{Residual gauge symmetries at null infinity} \label{resggescri}
We now discuss the null infinity limit of the residual gauge symmetries described in  \autoref{residualsec}. In order to do so, we will chose the functional form of the symmetry parameters that yields a finite and non-trivial action at null infinity. 

Recall that the symmetry parameters can be constructed from functions which depend only on the $y^\alpha=(V,Z)$ variables: $\Lambda(y)$  for the YM gauge symmetries and   $b(y)$ and $c(y)$ for the first and second family of gravitational symmetries. In terms of  Bondi coordinates \eqref{defbondicoords}, this translates into functions of the variables $(r,z)$. We will consider simple functional forms, $F(r,z)=r^n F(z)$, and choose $n$ so as to obtain well defined $r \to \infty $ limits. 

Let us start with the YM residual gauge symmetry \eqref{dellamPhi}. In order to have a well-defined action at null infinity, we need the RHS of \eqref{dellamPhi} to be $O(r^{-1})$. For the homogeneous term, this can only be achieved if $\Lambda=O(r^0)$. We are thus led to choose $\Lambda$ as
\be
\Lambda(r,z) = \Lambda_0(z) \label{defLam0}
\ee
Let us now discuss the inhomogeneous term in \eqref{dellamPhi}. To do so we note that the differential operator $S$ \eqref{defS} in Bondi coordinates takes the form
\be
 S \equiv \Omega_i^{\ \alpha}  x^i \partial_\alpha = - \zb u  \partial_u   - \zb^2 \partial_{\zb} +  \zb r \partial_r  + \frac{u}{r} \partial_z. \label{Oxdalpha}
\ee
When \eqref{Oxdalpha} acts on $\Lambda$ \eqref{defLam0}, only the last term survives, yielding a $O(r^{-1})$ result as expected. From here one can read off the action on the radiative data to be
\be
\delta_\Lambda \Phi_{\I} = u \partial_z \Lambda_0(z) +  i [\Lambda_0(z),\Phi_{\I}] . \label{delLamPhiscri}
\ee
The corresponding action on $A_z$ can be obtained from the relation \eqref{AzitoPhi}, where one recovers \eqref{delLambdaAz} for holomorphic $\Lambda_0$.

Consider now  the first family of gravitational symmetries, \eqref{delxiphi1}, with $b_\alpha = \partial_\alpha b$ as in Eq.\eqref{delxiphi1partial}. Writing the homogenous term in Bondi coordinates, one easily finds that, in order for this term to be $O(r^{-1})$ one needs to take $b=O(r)$. We then choose
\be
b(r,z) =-r f(z). \label{bitof}
\ee
The inhomogeneous term associated to $b$ turns out to be $O(r^{-1})$ due to subtle cancellations. One finally arrives at
\be
\delta_b \phi_{\I} = - u^2  f''(z)  + f(z) \partial_u \phi_{\I} \label{delfphi}
\ee
for the induced action on the free data. The corresponding action on  $C_{zz}$ can be obtained from the relation \eqref{Czzitophi}, where one recovers \eqref{delfCzz} for $f=f(z)$.
By similar considerations, one can extend the previous analysis to parameters $b_\alpha$ that are not necessarily a total derivative. One finds that, at null infinity, the only finite and non-trivial transformation that can be obtained is still of the form (\ref{delfphi}). Thus, the condition $b_\alpha = \partial_\alpha b$ does not imply a restriction as far as the asymptotic symmetries are concerned.

Let us finally discuss the second family of gravitational symmetries introduced in \autoref{2ndvfsec}. In this case one finds there are two possible choices of $c(y)$, either $O(r)$ or $O(r^2)$,  that yield a finite and non-trivial null infinity limit:
\be
c(r,z)  = r g(z)  \quad \text{or } \quad c(r,z) = - r^2 Y(z)/2  \label{citogY}
 \ee
For the first choice, one obtains
\be
\delta_c \phi_{\I} = -u^2 \zb  g''(z) + \zb g(z) \partial_u  \phi_{\I}, \label{delcgphi}
\ee
which, upon translating to $C_{zz}$, yields a supertranslation \eqref{delfCzz} with $f(z,\zb)= \zb g(z)$. 
For the second choice of $c$ in \eqref{citogY} one finds
\be \label{delcYphi}
\delta_c \phi_{\I} = -\frac{u^3}{6} Y'''(z) + (Y(z) \partial_z+ \frac{u}{2} Y'(z) \partial_u  +\frac{1}{2}Y'(z))  \phi_{\I},
\ee
which, when translated to $C_{zz}$ via \eqref{Czzitophi}, reproduces the superrotation action \eqref{delYCzz} with vector field \eqref{Yzbzero}.

\section{Color to kinematic symmetry map at null infinity} \label{DCsymscri}
Having understood how large gauge symmetries emerge from the residual symmetries discussed in  \autoref{residualsec}, we now study how the DC for `bulk'  symmetries described in   \autoref{Color to kinematics map for symmetries} translates into a DC for asymptotic symmetries at null infinity. This will be achieved by studying the DC symmetry relations to leading order in a $1/r$ expansion. Whereas for the first family (Eqs. \eqref{delLamsummary} and \eqref{dellamsummary}) this is rather straightforward, for the second family (Eqs. \eqref{delLamtsummary} and \eqref{dellamtsummary}) the expansion is trickier due to the appearance of the inverse wave operator. We will circumvent this difficulty by making use of the symmetry raising map discussed in \autoref{DCanddeltodeltsec}.

To get started, let us express the operators that appear in the description of bulk symmetries  in Bondi coordinates, paying special attention to their $1/r$ expansion. Given a spacetime function of the form $r^{n} F_n(u,z,\zb)$, we will regard $ F_n(u,z,\zb)$ as a function of ``weight" $n$ at null infinity ($n$ determines the the transformation properties of $ F_n(u,z,\zb)$ under a conformal rescaling).\footnote{Notice that this $n$ has nothing to do with the $n$ of \autoref{Double copy for an infinite family of symmetries}, which labeled the symmetry families.}

The PB in Bondi coordinates takes the form
\be \label{PB_with_r}
\{f,g\} = r^{-1} \{f,g\}_{\I} 
\ee
with
\be \label{PBscri}
\{f,g\}_{\I} =  (\partial_{\zb} f \partial_u g - \partial_u f \partial_{\zb} g ).
\ee
We will regard $\{ ,\}_{\I}$ as a PB defined intrinsically at null infinity (with weight $-1$).  Consider now the differential operator 
\be
S = \Omega_i^{\ \alpha}  x^i \partial_\alpha 
\ee
that features in the inhomogeneous part of the symmetries. Its form in Bondi coordinates was given in \eqref{Oxdalpha}. When acting on a function of the form $r^{n} F_n(u,z,\zb)$, it can be written as:
\be
S(r^{n} F_n) = r^n S_0(F_n) + r^{n-1} S_{-1}(F_n)
\ee
where
\ba
S_0 & = & - \zb u  \partial_u + n \zb - \zb^2 \partial_{\zb}  \\
 S_{-1} & = & u \partial_z.
\ea
$S_0$ and $S_{-1}$ will be regarded as differential operators defined intrinsically at null infinity. Note that  $S_0$ depends on the weight $n$ of the function being acted upon. 

We now proceed to describe the DC for the first and second family of asymptotic symmetries.

\subsection{First family} \label{First family}
The YM symmetry \eqref{delLamsummary} at null infinity was described in Eq. \eqref{delLamPhiscri} for $\Lambda = \Lambda_0(z)$. Using the notation introduced above, we write it as
\be
\delta_\Lambda \Phi_{\I} = S_{-1}(\Lambda_0) +  i [\Lambda_0,\Phi_{\I}] . \label{delLam0Phi}
\ee
For the gravity symmetry \eqref{dellamsummary}, we need to compute the `Hamiltonian' $\lambda$ \eqref{lamitob} for the choice of $b(y)$ given in \eqref{bitof}. This gives
\be
\lambda = 2 S( b=-r f(z))  =  r \lambda_1 + \lambda_0 \label{lambf}
\ee
with
\ba
\lambda_1  =-2 S_0(f) = - 2 \zb f(z) \label{lam1f}\\
 \lambda_0 = -2 S_{-1}(f)=  - 2 u f'(z). \label{lam0f}
\ea
The homogenous term in \eqref{dellamsummary} then has the expected order in $r$, since a $\lambda$ with an $O(r)$ piece will compensate for the $r^{-1}$ factor in the PB \eqref{PB_with_r}. Naively, it appears that the inhomogenous term could have $O(r)$ and $O(r^0)$ terms, thus spoiling the required $O(r^{-1})$ fall-offs. However, such coefficients turn out to be zero. In the end one finds  a well defined action on the radiative data, given by
\be
\delta \phi_{\I}   =  \frac{1}{2}S_{-1}(\lambda_0)   - \frac{1}{2} \{\lambda_1,\phi_{\I} \}_{\I} . \label{dellam0phi}
\ee
We emphasize that \eqref{dellam0phi} is just a rewriting of \eqref{delfphi}, but one that makes the DC structure manifest. Comparing \eqref{delLam0Phi} to \eqref{dellam0phi}, we find that the DC rules at null infinity have the same structure as those in the bulk, as expected. The new ingredient is that, in the homogeneous term, the weight of the symmetry parameter increases by 1 (this is to compensate for the factor of $1/r$ in the PB \eqref{PB_with_r}).

\subsection{Second family}
For the second family of symmetries, the DC relation in the bulk was made manifest by expressing the symmetries perturbatively,  Eqs. \eqref{delLamtsummary} and \eqref{dellamtsummary}.   Unfortunately, such expressions are difficult to expand in $1/r$ due to the  appearance of the inverse wave operator. However, the discussion given in  \autoref{DCanddeltodeltsec} offers an alternative route to obtaining the DC symmetry map: the second family of symmetries for each theory can be obtained from the first family via the  ``symmetry raising map'', see the vertical arrows of \eqref{diag1}. Thus, rather than going directly from $\delta_{\Lamt} \Phi$ to $\delta_{\lamt} \phi$ via the perturbative bulk DC map (bottom horizontal arrow of \eqref{diag1}), one can try to get there via the first family of symmetries. This is the strategy we will follow in this section to describe the second family of asymptotic symmetries and their DC relation.  


Let us start with a general discussion that applies to both theories; for concreteness we only display the equations corresponding to the YM case.  Since the first family of symmetries can be written as a sum of  $\O(\Phi^0)$ and $\O(\Phi^1)$ terms, $\delta \Phi = \delta^{(0)} \Phi + \delta^{(1)} \Phi$, the raising map \eqref{defdeltPhi} produces a symmetry $\delt$  that is at most $\O(\Phi^2)$: $\delt \Phi = \delt^{(0)} \Phi + \delt^{(1)} \Phi + \delt^{(2)} \Phi$.  Collecting powers of $\Phi$ in \eqref{defdeltPhi}, one finds each term should satisfy
\ba
\partial_i \delt^{(0)} \Phi & = &  \Omega_i^{\ \alpha} \partial_\alpha \delta^{(0)} \Phi  \label{delt0Phisec6},  \\ 
\partial_i \delt^{(1)} \Phi & = &  \Omega_i^{\ \alpha} \partial_\alpha \delta^{(1)} \Phi - i [\partial_i \Phi, \delta^{(0)} \Phi],  \label{delt1Phisec6} \\
\partial_i \delt^{(2)} \Phi & = &  - i [\partial_i \Phi, \delta^{(1)} \Phi] .  \label{delt2Phisec6}
\ea
Eq. \eqref{delt0Phisec6} was already discussed in  \autoref{DCanddeltodeltsec}, where it was found that 
\be \label{delt0Phisec7}
\delt^{(0)} \Phi = \frac{1}{2}S^2(\Lambda)  = \frac{1}{2} S(\Lamt), \quad \text{with} \quad \Lamt = \delta^{(0)}_\Lambda \Phi = S(\Lambda).
\ee
Similarly, if we start with a gravitational first family symmetry, so that $\delta^{(0)} \phi = \tfrac{1}{2} S(\lambda)$ (with $\lambda=2 S(b)$) one finds
\be \label{delt0phisec7}
\delt^{(0)} \phi = \frac{1}{6}S^2(\lambda)  = \frac{1}{3} S(\lamt), \quad \text{with} \quad \lamt = \delta^{(0)}_\lambda \phi = \frac{1}{2}S(\lambda).
\ee
The numerical factors in \eqref{delt0Phisec7} and \eqref{delt0phisec7} arise as particular cases of the general coefficients explained in \autoref{Double copy for an infinite family of symmetries}. We will later expand these expressions in $1/r$ in order to obtain $\delt^{(0)} \Phi_{\I}$ and $\delt^{(0)} \phi_{\I}$.  

To obtain $\delt^{(1)} \Phi_{\I}$ and $\delt^{(2)} \Phi_{\I}$, we will solve Eqs.  \eqref{delt1Phisec6} and \eqref{delt2Phisec6} asymptotically as follows. First, we note that the $x^i$ derivatives expressed in Bondi coordinates,
\ba
\partial_U & = & \partial_u \\
\partial_{\Zb} &= &  - z \partial_u + r^{-1}\partial_{\zb}
\ea
are not independent in the $r \to \infty$ limit. Thus, to leading order in $1/r$ it is enough to consider only one component of Eqs.  \eqref{delt1Phisec6} and \eqref{delt2Phisec6}. Choosing $i=U$ and writing them in Bondi coordinates one obtains
\ba
\partial_u \delt^{(1)} \Phi & = &   (- \zb \partial_u + r^{-1}\partial_{z}) \delta^{(1)} \Phi - i [\partial_u \Phi, \delta^{(0)} \Phi] \label{delt1Phiu}\\
\partial_u \delt^{(2)} \Phi & = &  - i [\partial_u \Phi, \delta^{(1)} \Phi] \label{delt2Phiu}
\ea
In order to have a well defined action at null infinity, we need the RHS of these equations to be $O(r^{-1})$. The resulting $O(r^{-1})$ factors will then give expressions for $\partial_u \delt^{(1)} \Phi_{\I}$ and $\partial_u \delt^{(2)} \Phi_{\I}$, from which the final answer can be obtained by  a single integral in $u$. 

Similar considerations apply to the gravitational case, in which case $\delt^{(1)} \phi_{\I}$, $ \delt^{(2)} \phi_{\I}$ are to be determined from the $O(r^{-1})$ part of the equations
\ba
\partial_u \delt^{(1)} \phi & = &   (- \zb \partial_u + r^{-1}\partial_{z}) \delta^{(1)} \phi +\tfrac{1}{2}\{\partial_u \phi, \delta^{(0)} \phi \} \label{delt1phiu}\\
\partial_u \delt^{(2)} \phi & = &  \tfrac{1}{2} \{ \partial_u \phi, \delta^{(1)} \phi \}. \label{delt2phiu}
\ea

In order to perform the required $1/r$ expansion, we need to specify the asymptotic  behavior of the  symmetry parameters $\Lambda$ and $\lambda$. We now describe two possibilities  that yield a well defined action of $\delt$ at null infinity. 

\subsubsection{$\delt$ from $\Lambda=O(r^{0})$,  $\lambda=O(r)$} \label{deltr0sec}
We start by considering symmetry parameters as those discussed in \autoref{First family},
\ba
\Lambda  & = & \Lambda_0(z)  \\
\lambda & = & r \lambda_1 + \lambda_0
\ea
with $\lambda$ as in Eqs. \eqref{lambf}, \eqref{lam1f}, \eqref{lam0f}.  The corresponding $\Lamt$ and $\lamt$ defined in Eqs. \eqref{delt0Phisec7} and \eqref{delt0phisec7} take the form
\be
\Lamt = r^{-1} \Lamt_{-1}, \quad  \lamt = r^{-1} \lamt_{-1}
\ee
with
\be
\Lamt_{-1}= S_{-1}(\Lambda_0) = u \Lambda_0'(z) 
\ee
\be
 \lamt_{-1} =  \tfrac{1}{2}S_{-1}(\lambda_0) = - u^2 f''(z).
\ee
One then obtains
\be
\delt^{(0)} \Phi_{\I}  = \frac{1}{2} S_{0}(\Lamt_{-1}) =  -  u \zb   \Lambda_0'(z)  \label{delt01Phi}
\ee
\be
\delt^{(0)} \phi_{\I} = \frac{1}{3}  S_{0}(\lamt_{-1}) =  u^2\zb  f''(z)
\ee
for the inhomogeneous part of the symmetry transformations. The linear part of the symmetries  can be obtained from Eqs. \eqref{delt1Phiu}, \eqref{delt1phiu}, taking into account the form of $\delta^{(0)}$ and $\delta^{(1)}$ given in \autoref{First family}. One  finds that the only terms contributing to order $O(r^{-1})$ are those coming from the $\zb \partial_u$ term hitting the commutator/PB. This allows for a trivial integration in $u$, leading to
\ba
\delt^{(1)} \Phi_{\I}  & = &  -i \zb  [\Lambda_0, \Phi_\I],   \label{delt11Phi} \\
\delt^{(1)} \phi_{\I} &  =&   \frac{1}{2}  \zb  \{\lambda_1,\phi_{\I}\}_{\I} .
\ea
Finally, one finds that the RHS of Eqs.  \eqref{delt2Phiu} and \eqref{delt2phiu}  decay faster than $O(r^{-1})$ and hence 
\be
\delt^{(2)} \Phi_{\I}   =  0 , \quad  \delt^{(2)} \phi_{\I} =  0.
\ee
The gravitational symmetry $\delt \phi_{\I}$ can be seen to coincide with a supertranslation associated to $\tilde{f}(z,\zb)=-\zb f(z)$  (Eq. \eqref{delcgphi} with $g(z) \to f(z)$). Thus, the symmetry raising map takes a supertranslation with $f=f(z)$ into a supertranslation  with $\tilde{f}=-\zb f(z)$.

\subsubsection{$\delt$ from $\Lambda=O(r)$, $\lambda=O(r^2)$ } \label{deltr1sec}
We now consider symmetry parameters $\Lambda$ and $\lambda$ of one order higher in $r$. Although these generate transformations that violate the $r \to \infty$ fall-offs of  the  YM and gravity fields,  the corresponding $\delt$ symmetries will turn out to preserve such fall-offs. 

On the YM side, we  consider a $O(r)$ gauge transformation 
\be
\Lambda = r \Lambda_1(z) + \Lambda_0  \label{OrLam}
\ee
where $\Lambda_0= u \partial_z \partial_{\zb} \Lambda_1(z) =0$ (as obtained from the condition $\square \Lambda=0$).  We explicitly keep this vanishing term in \eqref{OrLam} since it will get mapped into a non-trivial term under the DC.  
The corresponding $\Lamt$ is:
\be
\Lamt= S(\Lambda) = r S(\Lambda)_1  + S(\Lambda)_0 \label{SOrLam}
\ee
with\footnote{Since $S(\Lambda)$ has a non-trivial  $O(r^0)$ part, there is no need in this case to explicitly include the vanishing $\Lambda_0$ contribution.} 
\be
\Lamt_1=S(\Lambda)_1=S_0(\Lambda_1) = \zb \Lambda_1(z) , \quad \Lamt_0= S(\Lambda)_0 =  S_{-1}(\Lambda_1) = u \Lambda'_1(z).
\ee
Substituting this expansion in \eqref{delt0Phisec7} one finds\footnote{The structure of $\Lamt$  in this case is formally identical to the $\lambda$ of the first gravitational symmetry \eqref{lambf}, with $\Lambda_1 = -2 f$. The same cancellations occur when computing $S(\Lamt)$, so that only the $O(r^{-1})$ part survives.}
\be \label{delt0Phiscri}
\delt^{(0)} \Phi_{\I}  = \frac{1}{2} S_{-1}(\Lamt_0) =  \frac{u^2}{2} \Lambda''_1(z).
\ee
To obtain $\delt^{(1)} \Phi_{\I}$, we consider Eq. \eqref{delt1Phiu} for $\delta^{(0)}_\Lambda \Phi= S(\Lambda)$ and $\delta^{(1)}_\Lambda \Phi= i [\Lambda,\Phi]$.  Interestingly, the  potentially divergent terms cancel out, and one arrives at a finite equation for the null infinity free data $\Phi_{\I}$:
\be \label{delt1Phiscri}
\partial_u \delt^{(1)} \Phi_{\I} =  - i \zb \partial_u  [\Lambda_0, \Phi_{\I}] + i \partial_{z}[\Lambda_1, \Phi_{\I}]- i [\partial_u \Phi_{\I}, S_{-1}(\Lambda_1)]  .
\ee
Similarly, from  Eq. \eqref{delt2Phiu} one obtains
\be \label{delt2Phiscri}
\partial_u \delt^{(2)} \Phi_{\I}  =   [\partial_u \Phi_{\I} , [\Lambda_1,\Phi_{\I}]].
\ee
On the gravitational side,  we now consider a symmetry parameter $b$ that is $O(r^2)$ 
\be
b= - r^2 Y(z)/2. \label{bitoY}
\ee
The corresponding $\lambda=2 S(b)$ takes the form
\be
\lambda = r^2 \lambda_2 + r \lambda_1 \label{Or2lam}
\ee
with 
\be \label{lambda2ndii}
\lambda_2 = -2 \zb Y(z) , \quad \lambda_1 = -  u Y'(z).
\ee
$S(\lambda)=2 \lamt$ is then given by
\be
S(\lambda)  = r^2 S(\lambda)_2 +r S(\lambda)_1 + S(\lambda)_0  \label{SOr2lam}
\ee
with
\ba
S(\lambda)_2 & = & S_0(\lambda_2)  = -2 \zb^2 Y(z),  \\
S(\lambda)_1 &= &S_{-1}(\lambda_2) =  -2 u \zb Y'(z), \label{Sm1lam2} \\
S(\lambda)_0 & = &S_{-1}(\lambda_1)= -  u^2 Y''(z),
\ea
where we used that  $S_0(\lambda_1)=0$. To obtain $\delt^{(0)} \phi_{\I}$, we substitute the above expansion in \eqref{delt0phisec7}. It turns out that all potentially divergent terms cancel out and one is left with a $O(r^{-1})$ term, yielding
\be
\delt^{(0)} \phi_{\I}  = \frac{1}{3} S_{-1}(\lamt_0) =  -\frac{u^3}{6}Y'''(z).
\ee
This precisely reproduces the inhomogenous part of the superrotation action \eqref{delcYphi}.

To obtain $\delt^{(1)} \phi_{\I}$, we consider Eq. \eqref{delt1phiu} for $\delta^{(0)}_\lambda \phi= \tfrac{1}{2}S(\lambda)$ and $\delta^{(1)}_\lambda \phi= -\tfrac{1}{2} \{ \lambda,\phi\}$, with $\lambda$, $S(\lambda)$ as  given by \eqref{Or2lam},  \eqref{SOr2lam}. Once again the divergent terms cancel out, and one obtains a well defined equation at  null infinity:
\be \label{pudelt1phiscri}
\partial_u \delt^{(1)} \phi_{\I} =  \frac{1}{2}  \zb \partial_u  \{\lambda_1, \phi_{\I}\}_{\I} - \frac{1}{2} \partial_{z} \{\lambda_2, \phi_{\I}\}_{\I}+\frac{1}{4} \{\partial_u \phi_{\I},  S_{-1}(\lambda_2) \}_{\I}.  
\ee
This is the DC version of \eqref{delt1Phiscri}, with the rule described in \autoref{First family} for the increase in weight when going from $\Lambda$ to $\lambda$,  and the multiplicative $\mathfrak{r}$ factor associated to the operator $S$, see  \autoref{full_rules_summary}.  
Finally, the RHS of  \eqref{delt2Phiu} is found to be $O(r^{-2})$ and hence
\be
\delt^{(2)} \phi_{\I} =0.
\ee
It may be puzzling that the non-zero quadratic term \eqref{delt2Phiscri} trivializes after the DC. From the perspective of null infinity, this happens because the DC takes  the  weight $0$ commutator $[,]$ into the   weight $-1$ Poisson bracket $\{ , \}_{\I}$. Since in both cases one should have overall weight minus one, one would need to  map $\Lambda_1$ into a weight 3 $\lambda_3$ to get a non-trivial DC result. However there is no $\lambda_3$ in the gravitational symmetry under consideration.

We conclude by noting that if one expands \eqref{pudelt1phiscri} by explicitly writing the PB \eqref{PBscri} and the parameters  \eqref{lambda2ndii}, \eqref{Sm1lam2}, one recovers, as expected, the total $u$-derivative of the homogenous part of the superrotation action \eqref{delcYphi}. Thus, the symmetry raising map takes a ``divergent supertranslation'' defined by \eqref{bitoY} into a superrotation. 

\subsection{Summary}
\begin{figure}[H]
\centering
\includegraphics[scale=0.55]{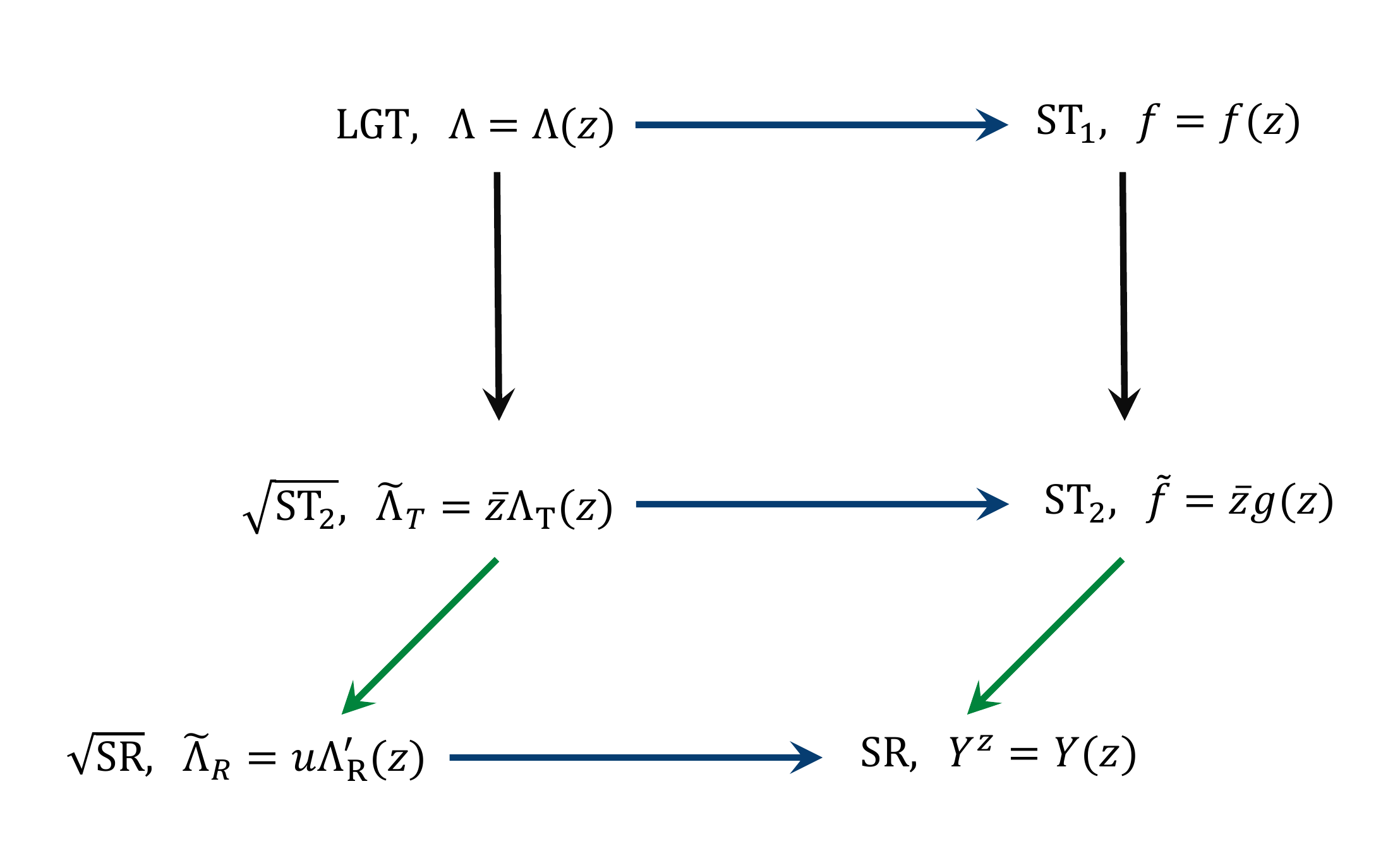}
\caption{Summary of the asymptotic symmetries and their various relations. LGT stands for non-abelian large gauge transformations, ST and SR for supertranslations and superrotations.  Horizontal lines represent the DC relations. Vertical lines represent the symmetry raising map between the first and second family of symmetries.  Diagonal arrows represent an increase in the power of $r$ of  symmetry parameters, see   main text for details.}
\label{fig:asymptsumm} 
\end{figure}

In \autoref{fig:asymptsumm} we summarize the different symmetries at null infinity, together with the functional form of their parameters. The diagram may be thought of as an augmented  version of \eqref{diag1}, with a new layer due to the two possibilities that arise in the second family of symmetries. In the first line we have holomorphic non-abelian large gauge transformations \eqref{delLamPhiscri} and holomorphic supertranslations \eqref{delfphi}. In the second line we have the non-gauge YM transformation \eqref{delt01Phi}, \eqref{delt11Phi} and the second type of allowed supertranslations \eqref{delcgphi}. The third line displays the non-gauge YM transformation defined by Eqs. \eqref{delt0Phiscri}, \eqref{delt1Phiscri}, \eqref{delt2Phiscri} and the superrotations \eqref{delcYphi}. 

The horizontal arrows represent the double copy relations between the different symmetries. The replacement rules are as in the bulk, with the following clarification: the commutator is replaced by the PB at null infinity  \eqref{PBscri}, which has weight $-1$. To compensate for this,  YM parameters of weight $n$ are replaced with Hamiltonians of weight $n+1$. 

Vertical arrows describe the symmetry raising map, see \autoref{DCanddeltodeltsec}. Diagonal arrows describe the increase in the power of $r$ associated to the parameter of the second family of symmetries, see e.g. Eq. \eqref{citogY} for the gravitational case.  To better understand these arrows, one can imagine completing  the diagram by  raising the power of $r$ in the parameters of the first family of YM and gravity symmetries, thus completing the cube. We can then  recover the bottom two symmetries by applying the symmetry raising map to these new vertices, see \autoref{deltr1sec} for details. We chose not to display them in the diagram since they do not preserve the $r \to \infty$ field fall-offs.\footnote{See the discussion section for further comments on this point.}

\section{Discussion} \label{Discussion}
In this paper we presented the double copy construction of a subset of  asymptotic symmetries. Our starting point was the study of residual symmetries of the self-dual sectors of YM and gravity, in the light-cone formulation. We identified two families of symmetries. For the first one, we found novel non-pertubative double copy rules in the bulk, leading, in the asymptotic regime, to the mapping of a holomorphic YM symmetry to a holomorphic supertranslation. The second family has a more striking structure, in the sense that an exact subset of diffeomorphism symmetries is obtained as the double copy of a perturbatively-defined and non-local transformation in the YM self-dual sector. At null infinity, we identify the YM origin of a subset of superrotations with a novel symmetry transformation, which to our knowledge has not been previously presented in the literature.

An important open problem is how to extend our DC symmetry prescription beyond the self-dual context. Promising avenues for extending the  Monteiro and O'Connell formulation have been given in e.g.  \cite{Fu:2016plh,Chen:2019ywi}. Another possibility would be to make use of the techniques developed in \cite{Ananth:2020ojp}, where the full BMS algebra is obtained from the residual symmetries of the gravitational theory in the light-cone gauge.  It would also be interesting to approach the problem via the formalism developed in \cite{Ashtekar:2020xll}. Alternatively, one could try to extend our analysis to other families of solutions for which the doubly copy is manifest. An obvious first step would be to look at solutions obtained from the Kerr-Schild ansatz \cite{Monteiro:2014cda}, which is structurally similar to the self-dual metric ansatz used in this paper.

The self-dual theories are known to possess infinitely many symmetries. An infinite tower of such symmetries can be obtained by iterating  the `symmetry raising map' reviewed in Eqs. \eqref{defdeltPhi}, \eqref{defdeltphi}. From this perspective, the two families of symmetries described above represent just the  first two  rungs of an infinite ladder.  Whereas we have demonstrated that DC relations in the bulk are obeyed at each level, we have not yet explored the realization of these higher level symmetries at null infinity. It is tempting to speculate that these  would be related, in the scattering amplitudes context, to the remarkable simplicity of  Maximally Helicity Violating (MHV) amplitudes. This expectation is justified by  existing explanations of MHV formulae based on the integrability properties of the self-dual equations, see e.g. \cite{Bardeen:1995gk,Rosly:1996vr,Mason:2008jy}. 

It is worthwhile to point out that, asymptotically, there is potentially yet another hierarchy if one allows for  gauge transformations that diverge as some power of $r$ \cite{Campiglia:2016hvg,Campiglia:2016efb,Campiglia:2018dyi,Seraj:2016jxi,Compere:2017wrj,Compere:2019odm}. There appears to be an interesting interplay between this hierarchy and the one described in the previous paragraph. Indeed, we were  lead to consider $O(r)$ divergent gauge symmetries in order to fully implement the raising map  between the first two families of asymptotic symmetries. It may be that the higher level families are related to gauge transformations diverging with higher powers of $r$.

\section{Aknowledgements}
We would like to thank Alok Laddha for illuminating discussions. MC acknowledges support from PEDECIBA and from ANII grant FCE-1-2019-1-155865. SN is supported by STFC grant ST/T000686/1. This research was supported by the Munich Institute for Astro- and Particle Physics (MIAPP) which is funded by the Deutsche Forschungsgemeinschaft (DFG, German Research Foundation) under Germany 's Excellence Strategy – EXC-2094 – 390783311.

\appendix

\section{Self-duality conditions} \label{sdapp}
\subsection{Yang-Mills field}
Given the field strength
\be
F_{\mu \nu} = \partial_\mu \A_\nu  - \partial_\nu \A_\mu - i [\A_\mu,\A_\mu],
\ee
its dual is defined as
\be
\tilde{F}_{\mu \nu} := \frac{1}{2}\epsilon_{\mu \nu}^{\phantom{\mu \nu }\rho \sigma} F_{\rho \sigma} . \label{dualF}
\ee
We say the field strength is self-dual if 
\be
\tilde{F}_{\mu \nu} = i F_{\mu \nu}. \label{selfdual}
\ee

In the notation of $(x^i,y^\alpha)$ coordinates, the independent components of the volume form $\epsilon_{\mu \nu \rho \sigma}$ are
\be
\epsilon_{\alpha \beta i j}=i \Pi_{\alpha \beta} \Omega_{i j}. \label{epsilonPiOmega}
\ee
The field strength of the YM field (\ref{Aphimu}) is given by
\ba
F_{ij} &= & 0 \\
F_{i \alpha} &= & \Pi_{\alpha}^{\ j} \partial_i \partial_j \Phi \\
F_{\alpha \beta} &= & 2 \Pi_{[\alpha}^{\ i }\partial_{\beta]}\partial_i \Phi - i \Pi_\alpha^{\ i} \Pi_\beta^{\ j} [\partial_i \Phi,\partial_j \Phi] \label{Falphabeta}
\ea
It is easy to see that $\tilde{F}_{i j}=0$.  Using (\ref{OmegaPi}) one can verify that $\tilde{F}_{i \alpha} = i F_{i \alpha}$. Let us then focus on $F_{\alpha \beta}$. Since it is antisymmetric in the 2d indices $(\alpha,\beta)$, it can be written as multiple of $\Pi_{\alpha \beta}$:
\be
F_{\alpha \beta} = f \Pi_{\alpha \beta} \quad \text{with} \quad f=- \frac{1}{2} \Omega^{\alpha \beta} F_{\alpha \beta}
\ee
The spacetime dual is then found to be given by $\tilde{F}_{\alpha \beta} =- i F_{\alpha \beta}$. Thus, the only way to satisfy the selfdual condition (\ref{selfdual}) is that this component vanishes:
\be
F_{\alpha \beta} =0 \iff \Omega^{\alpha \beta} F_{\alpha \beta}=0.
\ee
From (\ref{Falphabeta}) and using (\ref{OmegaPi}) one finds
\be
\Omega^{\alpha \beta} F_{\alpha \beta} = -( \square \Phi + i \Pi^{i j}[\partial_i \Phi,\partial_j \Phi] )
\ee
so that indeed it vanishes provided (\ref{eomPhi}) is satisfied. 
\subsection{Metric field}\label{App Metric field}
For spacetime metrics, the dual of the curvature tensor is defined as:
\be
\tilde{R}_{\mu \nu \rho}^{\phantom{\mu \nu \rho} \sigma} := \frac{1}{2}\epsilon_{\mu \nu}^{\phantom{\mu \nu }\eta \lambda} R_{\eta \lambda \rho}^{\phantom{\mu \nu \rho} \sigma},
\ee
and we say  the metric is self-dual if
\be
\tilde{R}_{\mu \nu \rho}^{\phantom{\mu \nu \rho} \sigma}  = i R_{\mu \nu \rho}^{\phantom{\mu \nu \rho} \sigma}. \label{dualR}
\ee
We now study this condition for the spacetime metric 
\be
g_{\mu \nu} = \eta_{\mu \nu}+h_{\mu \nu} \label{defg}
\ee
with  $h_{\mu \nu}$ given by \eqref{hphimu}.  Note that we are not requiring $h$ to be a perturbation, and regard (\ref{defg}) as a full, non-linear metric. The form of $\eta$ and $h$ imply the inverse metric is given exactly by 
\be
g^{\mu \nu} = \eta^{\mu \nu}-h^{\mu \nu} \label{inverseg}.
\ee
One can also check that the determinant of $g$ coincides with that of $\eta$. In particular, the volume form is still given by (\ref{epsilonPiOmega}).
The non-zero Christoffel  symbols are:
\be
\begin{split}
& \Gamma_{i \alpha}^j = \frac{1}{2} \eta^{\beta j}\partial_i h_{\alpha \beta}, \quad  \Gamma_{\alpha \beta}^i= \frac{1}{2} \eta^{\gamma i}(\partial_\alpha h_{\beta \gamma}+\partial_\beta h_{\alpha \gamma}- \partial_\gamma h_{\alpha \beta})+\frac{1}{2}h^{i j}\partial_j h_{\alpha \beta}, \\
& \Gamma_{\alpha \beta}^\gamma= - \frac{1}{2}\eta^{\gamma i}\partial_i h_{\alpha \beta}    
\end{split}
\ee
These expressions lead to the following independent components of the curvature tensor:
\ba
R_{i j k}^{\phantom{i j k} l} &=&0 \\
R_{i j k}^{\phantom{i j k} \alpha} &=& 0 \\
R_{i \alpha j}^{\phantom{i \alpha j} k} & =& -\frac{1}{2}\eta^{\beta k}\partial_i \partial_j h_{\alpha \beta} \\
R_{\alpha \beta \gamma}^{\phantom{\alpha \beta \gamma} \delta} & =& \frac{1}{2}\eta^{\delta i} \partial_{i}\partial_{\alpha} h_{\beta \gamma} + \frac{1}{4}\eta^{\delta i} \eta^{\epsilon j} \partial_j h_{\alpha \gamma} \partial_i h_{\beta \epsilon} - (\alpha \leftrightarrow \beta) \label{R3} \\
R_{\alpha \beta \gamma}^{\phantom{\alpha \beta \gamma} i} & =& \partial_\beta \Gamma_{\alpha \gamma}^i +\Gamma_{\alpha \gamma}^{\delta}\Gamma_{\delta \beta}^i+ \Gamma_{\alpha \gamma}^{j}\Gamma_{j \beta}^i - (\alpha \leftrightarrow \beta) \label{R4}
\ea
We now consider the dual curvature tensor (\ref{dualR}). Here it is important to note that the raising of indices of the volume form must be realized with the full inverse metric (\ref{inverseg}). We will however continue to raise and lower indices on $\Pi$ and $\Omega$ with the flat metric $\eta$.

It is easy to check that $\tilde{R}_{i j k}^{\phantom{i j k} l}=\tilde{R}_{i j k}^{\phantom{i j k} \alpha}=0$, as well as $\tilde{R}_{i \alpha j}^{\phantom{i \alpha j} k}=i R_{i \alpha j}^{\phantom{i \alpha j} k}$. We now discuss the self-dual condition for the components (\ref{R3}). Since the expression is antisymmetric in the 2d $\alpha,\beta$ indices, we can contract with $\Omega^{\alpha \beta}$. After some algebra, one finds
\be
\Omega^{\alpha \beta}(\tilde{R}_{\alpha \beta \gamma}^{\phantom{\alpha \beta \gamma} \delta} - i R_{\alpha \beta \gamma}^{\phantom{\alpha \beta \gamma} \delta})=-i \Pi_{\gamma}^{\ m} \eta^{\delta n }  \partial_m \partial_n \text{E}_\phi \label{sdgrav1}
\ee
with
\be
\text{E}_\phi=\square \phi - \frac{1}{2}\Pi^{i j} \Pi^{k l} \partial_i \partial_k \phi \partial_j \partial_l \phi
\ee
where $\text{E}_\phi=0$ is the SDE equation (\ref{eomphi}). Finally, the self-dual condition for (\ref{R4}), contracted with  $\Omega^{\alpha \beta}$,  gives
\be 
\Omega^{\alpha \beta}(\tilde{R}_{\alpha \beta \gamma}^{\phantom{\alpha \beta \gamma} i} - i R_{\alpha \beta \gamma}^{\phantom{\alpha \beta \gamma} i})=i\Pi^{ij}\partial_j\partial_\gamma\text{E}_\phi 
-i\eta^{i\alpha}\Pi_\gamma^{\ j}\partial_j\partial_\alpha \text{E}_\phi 
+i\Pi^{jk}\Pi_\gamma^{\ l}\Pi^{im}\partial_j\partial_l  \text{E}_\phi \partial_k\partial_m\phi . \label{sdgrav2}
\ee

\section{Perturbative transformations at higher orders} \label{perturbativeapp}
In a perturbative setting, the double copy rules \eqref{full_rules_summary} can be written as:
\be \label{full_rules_AppB}
\Phi^{(i)} \to \phi^{(i)},\quad  -i[ \ , \ ] \to \tfrac{1}{2} \{ \ , \ \},\quad \Lambda \to \lambda, \quad
\mathfrak{r}=\frac{\text{deg}(\Lambda)+1}{\text{deg}(\lambda)+1}
\ee
where
\be 
\begin{aligned}
\Phi =& \Phi^{(0)}+ \Phi^{(1)}+ \cdots \\
\phi =& \phi^{(0)}+ \phi^{(1)}+ \cdots
\end{aligned}
\ee

\subsection{Second order}
\subsubsection{First family}
The first family of symmetries was shown to double copy non-perturbatively in \autoref{First family of symmetries}. However, as a warm-up, we demonstrate the perturbative construction to second order in perturbation theory, before proceeding to the second family. Working to linear order, the first family of transformations acts on the YM scalar as (see \eqref{delphi0} and \eqref{delta_1_Phi_pert} ):
\be
\begin{aligned}
\delta_\Lambda \Phi^{(0)} = &  \Omega_j^{\ \beta} x^j \partial_\beta \Lambda,\\
\delta_\Lambda \Phi^{(1)} =& - i  [\Phi^{(0)},\Lambda]
\end{aligned}
\ee
We will now treat these as seeds and use the e.o.m. at second order in $\Phi$:
\be
\square \Phi^{(2)}=  - 2 i \Pi^{i j}[\partial_i \Phi^{(0)},\partial_j  \Phi^{(1)}]
\ee
to derive $ \delta_{\Lambda} \Phi^{(2)}$. We have
\be
\begin{aligned}
\square  \delta_{\Lambda}\Phi^{(2)}=&  - 2 i \Pi^{i j}[\partial_i   \delta_{\Lambda}\Phi^{(0)},\partial_j  \Phi^{(1)}]
- 2 i \Pi^{i j}[\partial_i \Phi^{(0)},\partial_j    \delta_{\Lambda}\Phi^{(1)}]\\
=&- 2 i \Pi^{i j}[ \Omega_i^{\ \beta}\partial_\beta \Lambda,\partial_j  \Phi^{(1)}]
- 2  \Pi^{i j}[\partial_i \Phi^{(0)}, [\partial_j \Phi^{(0)},\Lambda]]\\
=&i\square[\Lambda,\Phi^{(1)}]-i[\Lambda,\square\Phi^{(1)}]+\Pi^{ij}[\Lambda,[\partial_i\Phi^{(0)},\partial_j\Phi^{(0)}]]
\end{aligned}
\ee
where we used the Jacobi identity for the commutator and the fact that $\Lambda=\Lambda(y)$.  Then, using the eom for $\Phi^{(1)}$, we are left with
\be \label{YM_second_first}
\delta_\Lambda \Phi^{(2)} =- i  [\Phi^{(1)},\Lambda] 
\ee
Next, we look at the gravity transformation. Our seeds can be obtained by perturbing \eqref{delta_b_grav}:
\be
\begin{aligned}
\delta_\lambda \phi^{(0)} =& \tfrac{1}{2} \Omega_i^{\ \alpha} x^i \partial_\alpha \lambda \\
\delta_\lambda \phi^{(1)} =& \tfrac{1}{2}  \{\phi^{(0)},\lambda\}
\end{aligned}
\ee
We will now use the e.o.m. at second order in $\phi$,
\be
\square \phi^{(2)} =  \Pi^{i j} \Pi^{k l} \partial_i \partial_k \phi^{(0)} \partial_j \partial_l \phi^{(1)} 
\ee
to derive $ \delta_{\lambda} \phi^{(2)}$. We have
\be
\begin{aligned}
\square\delta_{\lambda}\phi^{(2)} =&  \tfrac{1}{2}\Pi^{i j} \Pi^{k l} \partial_i \partial_k \delta_{\lambda}\phi^{(0)} \partial_j \partial_l \phi^{(1)} 
+\tfrac{1}{2}\Pi^{i j} \Pi^{k l} \partial_i \partial_k \phi^{(0)} \partial_j \partial_l \delta_{\lambda}\phi^{(1)} \\
=& \Pi^{i j} \Pi^{k l} \partial_i \left(\Omega_k^{\ \alpha}\partial_\alpha\lambda \right) \partial_j \partial_l \phi^{(1)} 
+\tfrac{1}{2}\Pi^{i j} \Pi^{k l} \partial_i \partial_k \phi^{(0)} \partial_j \{\partial_l \phi^{(0)},\lambda\}\\
=&-\eta^{l\alpha}\Pi^{i j}\partial_i\partial_\alpha  \lambda \partial_j \partial_l \phi^{(1)}
-\tfrac{1}{4}\Pi^{kl}\{\lambda,\{\partial_k\phi^{(0)},\partial_l\phi^{(0)}\}\}\\
=&-\tfrac{1}{2}\square\{\lamt,\phi^{(1)}\}+\tfrac{1}{2}\{\lamt,\square\phi^{(1)}\} 
-\tfrac{1}{4}\Pi^{kl}\{\lamt,\{\partial_k\phi^{(0)},\partial_l\phi^{(0)}\}\}
\end{aligned}
\ee
where we used the Jacobi identity for the Poisson bracket and the fact that $\lambda$ is linear in $x$. Then, using the e.o.m. for $\phi^{(1)}$ we get
\be 
\delta_\lambda \phi^{(2)} =  \tfrac{1}{2} \{\phi^{(1)},\lambda\}
\ee
and, comparing with \eqref{YM_second_first}, we see that it is obtained from $\delta_\Lambda \Phi^{(2)}$ via the double copy rules \eqref{full_rules_AppB}.

\subsubsection{Second family}
Here we extend the construction of the second family of symmetries and their double copy to second order in perturbation theory. Working to linear order, the second family of transformations acts on the YM scalar as (see \eqref{deltaLamtPhi0} and \eqref{delLamtPhi1})
\be
\begin{aligned}
\delta_{\Lamt} \Phi^{(0)} =& \frac{1}{2}  \Omega_{i}^{\ \alpha}  x^i \partial_\alpha \Lamt ,\\
 \delta_{\Lamt} \Phi^{(1)}  =& - i  [ \Phi^{(0)}, \Lamt]  + 2 i \square^{-1} \eta^{i \alpha}  [\partial_\alpha \Phi^{(0)},\partial_i \Lamt] 
\end{aligned}
\ee
We will now treat these as seeds and use the e.o.m. at second order in $\Phi$:
\be
\square \Phi^{(2)}=  - 2 i \Pi^{i j}[\partial_i \Phi^{(0)},\partial_j  \Phi^{(1)}]
\ee
to derive $ \delta_{\Lamt} \Phi^{(2)}$. We have
\be
\begin{aligned}
\square  \delta_{\Lamt}\Phi^{(2)}=&  - 2 i \Pi^{i j}[\partial_i   \delta_{\Lamt}\Phi^{(0)},\partial_j  \Phi^{(1)}]
- 2 i \Pi^{i j}[\partial_i \Phi^{(0)},\partial_j    \delta_{\Lamt}\Phi^{(1)}]\\
=& - 2 i \Pi^{i j}[  \Omega_{i}^{\ \alpha} \partial_\alpha \Lamt,\partial_j  \Phi^{(1)}]
-2\Pi^{ij}[\partial_i \Phi^{(0)}, [\partial_j\Phi^{(0)}, \Lamt]]-2\Pi^{ij}[\partial_i \Phi^{(0)}, [\Phi^{(0)},\partial_j\Lamt]]\\
&+4\Pi^{ij}[\partial_i \Phi^{(0)},\partial_j  \square^{-1} \eta^{k \alpha}  [\partial_\alpha \Phi^{(0)},\partial_k \Lamt]] \\
=&2i\eta^{i\alpha}[\partial_\alpha\Lamt,\partial_i  \Phi^{(1)}]+\Pi^{ij}[\Lamt,[\partial_i\Phi^{(0)},\partial_j\Phi^{(0)}]]
-2\Pi^{ij}[\partial_i \Phi^{(0)}, [\Phi^{(0)},\partial_j\Lamt]]\\
&+4\Pi^{ij}[\partial_i \Phi^{(0)},\partial_j  \square^{-1} \eta^{k \alpha}  [\partial_\alpha \Phi^{(0)},\partial_k \Lamt]] \\
=&i\square [\Lamt,\Phi^{(1)}]-2i\eta^{i\alpha}[\partial_i \Lamt,\partial_\alpha\Phi^{(1)}]-i[\Lamt,\square\Phi^{(1)}]
+\Pi^{ij}[\Lamt,[\partial_i\Phi^{(0)},\partial_j\Phi^{(0)}]]\\
&-2\Pi^{ij}[\partial_i \Phi^{(0)}, [\Phi^{(0)},\partial_j\Lamt]]
+4\Pi^{ij}[\partial_i \Phi^{(0)},\partial_j  \square^{-1} \eta^{k \alpha}  [\partial_\alpha \Phi^{(0)},\partial_k \Lamt]] 
\end{aligned} 
\ee 
where we used the Jacobi identity to get to the third line and $\square\Lamt=0$ to get to the fourth line. Then, using the eom for $\Phi^{(1)}$, we are left with
\be 
\label{YM_second_second}
\begin{aligned}
 \delta_{\Lamt} \Phi^{(2)}  =& - i  [ \Phi^{(1)}, \Lamt]  + 2 i \square^{-1} \eta^{i \alpha}  [\partial_\alpha \Phi^{(1)},\partial_i \Lamt] \\
&-2\square^{-1}\Pi^{ij}[\partial_i \Phi^{(0)}, [\Phi^{(0)},\partial_j\Lamt]]
+4\square^{-1}\Pi^{ij}[\partial_i \Phi^{(0)},\partial_j  \square^{-1} \eta^{k \alpha}  [\partial_\alpha \Phi^{(0)},\partial_k \Lamt]]
\end{aligned}
\ee
Next, we look at the gravity transformation. Our seeds are \eqref{dellamtsummary}:
\be
\begin{aligned}
\delta_{\lamt} \phi^{(0)} =& \frac{1}{3}  \Omega_{i}^{\ \alpha}  x^i \partial_\alpha \lamt \\
\delta_{\lamt} \phi^{(1)} =&\tfrac{1}{2} \{ \phi^{(0)}, \lamt \}  -  \square^{-1} \eta^{i \alpha} \{ \partial_\alpha \phi^{(0)}, \partial_i \lamt \}
\end{aligned}
\ee
We will now use the e.o.m. at second order in $\phi$,
\be
\square \phi^{(2)} =  \Pi^{i j} \Pi^{k l} \partial_i \partial_k \phi^{(0)} \partial_j \partial_l \phi^{(1)} 
\ee
to derive $ \delta_{\lamt} \phi^{(2)}$. We have
\be
\begin{aligned}
\square\delta_{\lamt}\phi^{(2)} =& \tfrac{1}{2} \Pi^{i j} \Pi^{k l} \partial_i \partial_k \delta_{\lamt}\phi^{(0)} \partial_j \partial_l \phi^{(1)} 
+\tfrac{1}{2} \Pi^{i j} \Pi^{k l} \partial_i \partial_k \phi^{(0)} \partial_j \partial_l \delta_{\lamt}\phi^{(1)} \\
=&\Pi^{i j} \Pi^{k l} \partial_i \left(\Omega_k^{\ \alpha}\partial_\alpha  \lamt\right)  \partial_j \partial_l \phi^{(1)}
+\tfrac{1}{2} \Pi^{i j} \Pi^{k l} \partial_i \partial_k \phi^{(0)} \partial_j  \{\partial_l  \phi^{(0)}, \lamt \} \\
&+\tfrac{1}{2} \Pi^{i j} \Pi^{k l} \partial_i \partial_k \phi^{(0)} \partial_j \{ \phi^{(0)},  \partial_l \lamt \} 
-\Pi^{i j} \Pi^{k l} \partial_i \partial_k \phi^{(0)} \partial_j \partial_l \square^{-1} \eta^{i \alpha} \{ \partial_\alpha \phi^{(0)}, \partial_i \lamt \}\\
=&-\eta^{l\alpha}\Pi^{i j}\partial_i\partial_\alpha  \lamt \partial_j \partial_l \phi^{(1)}
-\tfrac{1}{4}\Pi^{kl}\{\lamt,\{\partial_k\phi^{(0)},\partial_l\phi^{(0)}\}\}\\
&+\tfrac{1}{2} \Pi^{i j} \Pi^{k l} \partial_i \partial_k \phi^{(0)} \partial_j \{ \phi^{(0)},  \partial_l \lamt \} 
-\Pi^{i j} \Pi^{k l} \partial_i \partial_k \phi^{(0)} \partial_j \partial_l \square^{-1} \eta^{i \alpha} \{ \partial_\alpha \phi^{(0)}, \partial_i \lamt \}\\
=&-\tfrac{1}{2} \square\{\lamt,\phi^{(1)}\}+\eta^{l\alpha}\{\partial_l  \lamt,\partial_\alpha \phi^{(1)}\}
+\tfrac{1}{2} \{\lamt,\square\phi^{(1)}\}-\tfrac{1}{4}\Pi^{kl}\{\lamt,\{\partial_k\phi^{(0)},\partial_l\phi^{(0)}\}\}\\
&+\tfrac{1}{2} \Pi^{i j} \Pi^{k l} \partial_i \partial_k \phi^{(0)} \partial_j \{ \phi^{(0)},  \partial_l \lamt \} 
-\Pi^{i j} \Pi^{k l} \partial_i \partial_k \phi^{(0)} \partial_j \partial_l \square^{-1} \eta^{i \alpha} \{ \partial_\alpha \phi^{(0)}, \partial_i \lamt \}\\
\end{aligned}
\ee
where we used the Jacobi identity for the Poisson bracket to get to the third line and $\square\lamt=0$ to get to the fourth line. Then, using the e.o.m. for $\phi^{(1)}$, we get
\be
\label{grav_delta_2}
\begin{aligned}
\delta_{\lamt}\phi^{(2)}=&   \tfrac{1}{2} \{ \phi^{(1)}, \lamt \}  -  \square^{-1} \eta^{i \alpha} \{ \partial_\alpha \phi^{(1)}, \partial_i \lamt \}\\
&+\tfrac{1}{2}  \square^{-1}\Pi^{ij} \{ \partial_i \phi^{(0)},\{ \phi^{(0)},  \partial_j \lamt \}\} 
-\square^{-1}\Pi^{i j}  \{\partial_i  \phi^{(0)}, \partial_j  \square^{-1} \eta^{i \alpha} \{ \partial_\alpha \phi^{(0)}, \partial_i \lamt \}\}
\end{aligned} 
\ee 
and comparing with the YM transformation \eqref{YM_second_second} we see that it double copies correctly under the rules \eqref{full_rules_AppB}.

\subsection{Recursive construction at arbitrary orders} \label{Recursive construction at arbitrary orders}
We will demonstrate recursively that the double copy rules \eqref{full_rules_AppB} work at all orders in perturbation theory. The proof applies simultaneously to the first and second family of symmetries (and, indeed, to the infinite set of families constructed in \autoref{Double copy for an infinite family of symmetries}). Start by assuming we have shown that the transformation rules for self-dual YM up to some order $(n-1)$:
\be 
\delta_{\Lambda} \Phi^{(0)},\delta_{\Lambda} \Phi^{(1)},\cdots,\delta_{\Lambda}\Phi^{(n-1)} \label{YMtranfsn_1}
\ee
double copy correctly, under the rules \eqref{full_rules_AppB}, into the respective gravity transformations:
\be 
\label{grav_seeds_n_1}
\delta_{\lambda} \phi^{(0)},\delta_{\lambda} \phi^{(1)},\cdots,\delta_{\lambda}\phi^{(n-1)} 
\ee
We now treat the transformations $\delta_{\Lamt} \Phi^{(0)},\cdots,\delta_{\Lamt}\Phi^{(n-1)}$ as seeds, and make use of the YM self-dual equation at order $n$:
\be 
\square \Phi^{(n)} =  - i \Pi^{i j}\sum_{m+p=n-1}[\partial_i \Phi^{(m)},\partial_j \Phi^{(p)}] 
\ee
to derive
\be
\delta_{\Lamt} \Phi^{(n)} =  - i \Pi^{i j}\square^{-1} \sum_{m+p=n-1}[\partial_i \delta_{\Lamt} \Phi^{(m)},\partial_j \Phi^{(p)}]  
 - i \Pi^{i j}\square^{-1} \sum_{m+p=n-1}[\partial_i \Phi^{(m)},\partial_j \delta_{\Lamt} \Phi^{(p)}]  
\ee
Then using the d.c. rules \eqref{full_rules_AppB} and the assumption that the double copy works up to order $(n-1)$ we get
\be
\delta_{\lamt} \phi^{(n)} =  \tfrac{1}{2} \Pi^{i j}\square^{-1} \sum_{m+p=n-1}\{\partial_i \delta_{\lamt} \phi^{(m)},\partial_j \phi^{(p)}\}  
 +\tfrac{1}{2} \Pi^{i j}\square^{-1} \sum_{m+p=n-1}\{\partial_i \phi^{(m)},\partial_j \delta_{\lamt} \phi^{(p)}\}
\ee
Separately, we can directly derive the gravity transformation by taking $\delta_{\lamt} \phi^{(0)},\cdots,\delta_{\lamt} \phi^{(n-1)}$ as seeds, and using the gravity self-dual equation at order n:
\be 
\square \phi^{(n)}=\tfrac{1}{2}\Pi^{ij}\sum_{m+p=n-1}\{\partial_i \phi^{(m)},\partial_j \phi^{(p)}\} 
\ee
we get
\be 
\square \delta_{\lamt}\phi^{(n)}=\tfrac{1}{2}\Pi^{ij}\sum_{m+p=n-1}\{\partial_i \delta_{\lamt}\phi^{(m)},\partial_j \phi^{(p)}\} 
+\tfrac{1}{2}\Pi^{ij}\sum_{m+p=n-1}\{\partial_i \phi^{(m)},\partial_j \delta_{\lamt}\phi^{(p)}\} 
\ee
so indeed
\be 
\delta_{\lamt} \phi^{(n)} =  \tfrac{1}{2} \Pi^{i j}\square^{-1} \sum_{m+p=n-1}\{\partial_i \delta_{\lamt} \phi^{(m)},\partial_j \phi^{(p)}\}  
 +\tfrac{1}{2} \Pi^{i j}\square^{-1} \sum_{m+p=n-1}\{\partial_i \phi^{(m)},\partial_j \delta_{\lamt} \phi^{(p)}\}
\ee
as needed. As seen in this section, the fact that the DC rules continue to hold at all orders can be traced back to the fact that the e.o.m. themselves exhibit a DC structure.

\section{Comparison with convolution DC} \label{convoapp}
At zeroth order in the fields there exists an alternative DC construction for gravity and YM symmetries, based on a convolution dictionary \cite{Anastasiou:2014qba,Anastasiou:2018rdx,LopesCardoso:2018xes}.  In this appendix we will show that our DC is consistent with the convolution one where they overlap. Since the latter relates linearized YM and gravity gauge symmetries, we will compare it with our first family at zeroth order in the fields.

For simplicity, we present the comparison in the case of $(2,2)$ signature,\footnote{See e.g. \cite{Atanasov:2021oyu} for a recent discussion.} where $Z$ and $\Zb$ are treated as independent real variables. This will allow us to perform Fourier transforms along $Z$ and $\Zb$ independently.

In momentum space, the convolution DC relates a linearized gravity diffeomorphism $\tilde{\xi}_\mu(p)$ to a linearized YM gauge parameter $\tilde{\Lambda}(p)$ by\footnote{Note $\tilde{\Lambda}$ is the Fourier transform of the gauge parameter $\Lambda$. It should not be confused with the parameter of the second family of symmetries, which in the main text is also denoted by $\tilde{\Lambda}$. For simplicity we have assumed that the two gauge sectors are identical.}
\be
\tilde{\xi}_\mu(p) = \Tr[ \tilde{\Lambda}(p)  \tilde{\varphi}^{-1}(p) \tilde{\A}_\mu(p)  ] \label{convxilam}
\ee
where $\varphi$ is the biadjoint ``spectactor'' scalar field. Since $A_i=0$, we immediately see that 
\be
\xi_i=0,
\ee
consistent with \eqref{xiizero}. Next we note that since $\Lambda=\Lambda(y)$ is independent of $x^i=(U,\Zb)$ it follows that
\be
\Lamt(p) = \delta(p_U) \delta(p_{\Zb}) \Lamt(p_V,p_Z) .
\ee
Substituting in \eqref{convxilam}, we find that $\tilde{\xi}_\mu$ is proportional to $ \delta(p_U) \delta(p_{\Zb})$. Hence, when transformed back to position space, it can only depend on the $y$ variables
\be
\xi_\alpha = \xi_\alpha(y), 
\ee
consistent with \eqref{xiizero}.

\bibliography{Ref_Lib_asympt_v2}
\bibliographystyle{utphys}


\end{document}